\documentclass[useAMS,usenatbib, letterpaper]{mn2e}
\usepackage{graphicx}
\usepackage{grffile}
\usepackage{amssymb}
\usepackage{textcomp}
\usepackage{color}
\usepackage{xcolor}
\usepackage{natbib}
\usepackage{appendix}
\usepackage{amsmath}
\usepackage{morefloats}
\usepackage{subfigure}
\usepackage{multirow}
\usepackage{threeparttable,upgreek}
\usepackage{glossaries}

\newacronym{AGB}{AGB}{asymptotic giant branch}
\newacronym{BOSS}{BOSS}{Baryon Oscillation Spectroscopic Survey}
\newacronym{CSP}{CSP}{composite stellar population}
\newacronym{DES}{DES}{Dark Energy Survey}
\newacronym{GALEX}{GALEX}{Galaxy Evolution Explorer}
\newacronym{FIR}{FIR}{far infrared}
\newacronym{FUV}{FUV}{far ultraviolet}
\newacronym{IMF}{IMF}{initial mass function}
\newacronym{IR}{IR}{infrared}
\newacronym[longplural={luminous red galaxies}]{LRG}{LRG}{luminous red galaxy}
\newacronym{NIR}{NIR}{near infrared}
\newacronym{NUV}{NUV}{near ultraviolet}
\newacronym{panstarrs}{Pan-STARRS}{Panoramic Survey Telescope and Rapid Response System}
\newacronym[longplural = {principal component analyses}]{PCA}{PCA}{principal component analysis}
\newacronym{PDF}{PDF}{probability distribution function}
\newacronym{photoz}{photo-$z$}{photometric redshift}
\newacronym{SDSS}{SDSS}{Sloan Digital Sky Survey}
\newacronym{SED}{SED}{spectral energy distribution}
\newacronym{SFH}{SFH}{star formation history}
\newacronym{SFR}{SFR}{star formation rate}
\newacronym{SOM}{SOM}{self-organising map}
\newacronym{specz}{spec-$z$}{spectroscopic redshift}
\newacronym{SSP}{SSP}{single stellar population}
\newacronym{UV}{UV}{ultraviolet}

\newcommand{\sz}{spec-$z$}

\definecolor{colb}{HTML}{4169E1}
\definecolor{colg}{HTML}{00CD66}
\definecolor{colr}{HTML}{CD2626}
\definecolor{pink}{HTML}{CD2990}

\newcommand{\ttt}{\texttt}

\newcommand{\mzs}{z_{\mathrm{spec}}}
\newcommand{\mzp}{z_{\mathrm{phot}}}
\newcommand{\mAng}{{\text{\AA}}}

\newcommand{\Gyr}{\,\mathrm{Gyr}}

\newcommand{\mli}{\lambda_i}

\newcommand{\dz}{\Delta z}
\newcommand{\dzrf}{\Delta z_\mathit{rf}}

\newcommand{\sdzrf}{\sigma(\Delta z_\mathit{rf})}
\newcommand{\ssixtyeight}{\sigma_{68}}
\newcommand{\ssixtyeightrf}{\sigma_{68,\mathit{rf}}}
\newcommand{\snmad}{\sigma_\mathrm{NMAD}}
\newcommand{\stwosrf}{\sigma_{95,\mathit{rf}}}
\newcommand{\sthreesrf}{\sigma_{99,\mathit{rf}}}

\newcommand{\cezgal}{\texttt{EzGal}}
\newcommand{\cphotoz}{\texttt{PhotoZ}}
\newcommand{\csedfit}{\texttt{SEDfit}}

\newcommand{\cgalaxev}{\texttt{GALAXEV}}

\newcommand{\bcexpb}{BC03 $\lambda^{-\beta}$}
\newcommand{\expb}{$\lambda^{-\beta}$}
\newcommand{\Ang}{\text{\AA}}
\newcommand{\br}{4000\,\AA{} break}

\newcommand{\umg}{$u-g$}
\newcommand{\gmr}{$g-r$}
\newcommand{\rmi}{$r-i$}
\newcommand{\imz}{$i-z$}

\newcommand{\mZsun}{Z_{\odot} }

\title[Photo-zs and Model SEDs of BOSS CMASS Galaxies]{Photometric Redshifts and Model Spectral Energy Distributions of Galaxies From the SDSS-III BOSS DR10 Data}
\author[N. Greisel et al.]{\parbox{\textwidth}{N. Greisel$^{1,2,3}$\thanks{e-mail:ngreisel@cefca.es}, S. Seitz$^{1,2}$, N. Drory$^{2,4}$, R. Bender$^{1,2}$, R.~P. Saglia$^{1,2}$, J. Snigula$^{1,2}$}\vspace{0.4cm}\\
\parbox{\textwidth}{$^{1}$University Observatory Munich, Scheinerstrasse 1, 81679 Munich, Germany\\
$^{2}$Max Planck Institute for Extraterrestrial Physics, Giessenbachstrasse, 85748 Garching, Germany\\
$^{3}$Centro de Estudios de F\'{\i}sica del Cosmos de Arag\'{o}n (CEFCA), Plaza San Juan 1, 44001 Teruel, Spain\\
$^{4}$Department of Astronomy, The University of Texas at Austin, 2515 Speedway, Stop C1400, Austin, TX 78712, USA}}
\begin{document}
\glsunset{BOSS}
\glsunset{DES}
\glsunset{GALEX}
\glsunset{panstarrs}
\glsunset{SDSS}
\glsunset{SOM}
\date{}
\maketitle
\label{firstpage}
\begin{abstract}
We construct a set of model spectra specifically designed to match the colours of the BOSS CMASS galaxies and to be used with photometric redshift template fitting techniques.
As a basis we use a set of \glspl{SED} of single and composite stellar population models.
These models cannot describe well the whole colour range populated by the CMASS galaxies at all redshifts, wherefore we modify them by multiplying the \glspl{SED} with $\lambda^{-\beta}$ for $\lambda>\lambda_i$ for different values of $\lambda_i$ and $\beta$.
When fitting these \glspl{SED} to the colours of the CMASS sample, with a burst and dust components in superposition, we can recreate the location in colour spaces inhabited by the CMASS galaxies.
From the best fitting models we select a small subset in a two-dimensional plane, whereto the galaxies were mapped by a self-organizing map.
These models are used for the estimation of photometric redshifts with a Bayesian template fitting code.
The photometric redshifts with the novel templates have a very small outlier rate of $0.22\,\%$, a low bias $\langle\Delta z/(1+z)\rangle=2.0\cdot10^{-3}$, and scatter of $\ssixtyeight=0.026$ in the restframe.
Using our models, the galaxy colours are reproduced to a better extent with the photometric redshifts of this work than with photometric redshifts of SDSS.
\end{abstract}


\glsresetall
\glsunset{BOSS}
\glsunset{DES}
\glsunset{GALEX}
\glsunset{panstarrs}
\glsunset{SDSS}

\section{Introduction}
Spectroscopic surveys provide very precise measurements of the cosmological redshift, but they are time consuming and cannot be applied to fainter galaxies.
Significantly larger volumes of the cosmos can be probed on shorter timescales with photometric surveys by observation through selected filter bands.
Since spectral features cannot be resolved by medium- or broadband photometry, one has to apply statistical methods to derive the \gls{photoz}.
Redshift measurements are necessary in many cosmological contexts whenever information about the redshift tomography, or the distribution of galaxies as a function of redshift is required.
The results of cosmological applications strongly depend on the photometric redshift accuracies.
Large biases and scatter in the \glspl{photoz} can deteriorate any such study, e.g., the dark energy constraints from shear tomography \citep{2006ApJ...636...21M}, or the baryonic acoustic oscillation scale \citep{2009ApJ...691..241B,2011MNRAS.411..277S}.
However, statistical errors can be accounted for if they are well known \citep[e.g.,][]{2006MNRAS.366..101H}.
Furthermore, in opposition to a single \gls{photoz} estimate, including the probability density function (PDF) $P(z)$ in the analysis of the \gls{photoz} uncertainties enhances the accuracy of cosmological measurements \citep[e.g.,][]{2008MNRAS.386..781M,2012MNRAS.421.2355H}.
\\
The techniques for \gls{photoz} estimation are commonly divided into two categories: template fitting and empirical methods.
Empirical methods learn a relation between the photometric observables and the spectroscopic redshift (spec-$z$) of a training set of galaxies in order to employ that relation to objects without spectroscopic information \citep[e.g.,][]{2004PASP..116..345C,2009AAS...21348303G,2013MNRAS.432.1483C,2014MNRAS.438.3409C}.
These techniques have the great advantage that they take calibration errors explicitly into account, and that they can include photometric observables other than magnitudes and colours.
Nevertheless, they strongly rely on the training sample which has to be a good representation of the query galaxies to yield accurate \glspl{photoz}.
Furthermore, empirical methods have to be trained anew when they are applied to different surveys, filter systems, or extracted magnitudes or fluxes.
Also, many empirical methods do not take photometric measurement uncertainties into account.\\
In contrast to empirical methods, template fitting techniques essentially perform a maximum likelihood fit to the data \citep[e.g.,][]{1999MNRAS.310..540A,2000ApJ...536..571B,2000A&A...363..476B,2006A&A...457..841I,2006MNRAS.372..565F,2008ApJ...686.1503B}.
A set of template \glspl{SED} are shifted to several redshift steps, where they are multiplied with the filter functions and integrated over.
The such predicted fluxes of the templates are then fitted to the photometry of the data.
Therefore, template fitting techniques can be applied to data from any photometric system, and, concurrently to the \gls{photoz} estimation, also provide restframe properties.
In order to predict the \glspl{photoz} to a high precision, the data has to be well calibrated (or a catalogue with precise redshift measurements has to be available to re-calibrate the zeropoints).
Furthermore, it is of paramount importance that the underlying model \glspl{SED} represent the data in question.
In order to achieve that, some codes use combinations of model \glspl{SED}, or ``repair'' the templates \citep[e.g.,][]{2003AJ....125..580C,2006MNRAS.372..565F,2008ApJ...686.1503B}.
However, the underlying \glspl{SED} have to be already well selected if a modification of them should succeed as a representation of the data.
Moreover, if one introduces templates that do not match the data, the quality of the photometric redshifts is deteriorated.
Therefore, we generate in this work a model set designed to match the data in question.\\
In \citet[][hereafter G13]{2013ApJ...768..117G} we created model \glspl{SED} for spectroscopically observed \glspl{LRG} from the Sloan Digital Sky Survey data release 7 \citep[SDSS-II,][]{2000AJ....120.1579Y,2001AJ....122.2267E,2009ApJS..182..543A}.
These models were then used for the estimation of \glspl{photoz} and yielded accurate results.
The \gls{LRG} sample of \gls{SDSS}-II included \glspl{LRG} only up to redshift $z\approx0.5$.
In this paper we extend the work done in G13 and generate a set of model \glspl{SED} on the basis of the CMASS sample of the \emph{Baryon Oscillation Spectroscopic Survey} \citep[BOSS,][]{2011AJ....142...72E,2013AJ....145...10D,2014ApJS..211...17A}.
These templates are created on the basis of the colours of the CMASS galaxies.
They are specifically designed and selected to produce accurate photometric redshifts with template fitting techniques.
\\
This paper is organized as follows.
We present the template fitting photometric redshift code used in this work in Sec.~\ref{sec:paper2PhotoZ}.
In Sec.~\ref{sec:paper2BOSSCMASSsample} the data used in this work is described.
We briefly explain the \gls{SED} fitting routine in Sec.~\ref{sec:paper2newsedtemplates}.
After that, we will go into detail about the generation of model \glspl{SED} and how we modify them in order to fit the data colours to a better extent.
At the end of Sec.~\ref{sec:paper2newsedtemplates} we select a model set which is to be used with template fitting photometric redshift codes.
In Sec.~\ref{sec:BOSSphotozresults} we present the photometric redshift results of our new models and compare them with the two different \glspl{photoz} available on the \gls{SDSS} database.
We also predict the colours of our models when fitted to the more accurate one of the \gls{SDSS} \glspl{photoz} in comparison to their prediction with the redshifts of this work.
Finally, we give a summary and conclusions in Sec.~\ref{sec:paper2summary}.\\
Throughout this paper we assume a $\Lambda$CDM cosmology with $\Omega_m=0.3$, $\Omega_\Lambda=0.7$ and $H_0=70\,\mathrm{km\,s}^{-1}\mathrm{Mpc}^{-1}$.
Magnitudes are $\log_{10}$ magnitudes given in the AB system \citep{1983ApJ...266..713O}.


\section{Photometric Redshifts}
\label{sec:paper2PhotoZ}
For photometric redshift estimation in this paper we use the Bayesian template fitting code \cphotoz\ \citep{2001defi.conf...96B}.
Template fitting codes essentially determine the photometric redshift by performing a maximum likelihood fit of the predicted colours of a set of template \glspl{SED}, calculated by the multiplication of the \glspl{SED} with the survey filter functions and integrated, to the observed galaxy colours at varying redshifts.
In order to lift degeneracies in the colours of model \glspl{SED}, Bayesian codes include the possibility to assign prior probabilities in redshift and luminosity to individual models.
The resulting probability of a model-redshift combination then reads
\begin{equation}
  P(\boldsymbol{\mu}|C,m)\propto \mathcal{L}(C,m|\boldsymbol{\mu})\cdot P(\boldsymbol{\mu}),\nonumber
\end{equation}
where $C$ and $m$ are the colours and magnitudes of the photometric data, while $\boldsymbol{\mu}$ denotes the absolute magnitude $M$ and the redshift $z$.
The second factor is the prior probability $P(\boldsymbol{\mu}|T)=P_M(M|T)\cdot P_z(z|T)$, with the \glspl{PDF} of $M$ and $z$ for a template $T$.
In the case of the \cphotoz\ code, the prior functions follow
\begin{equation}
  P_x(x|T)\propto\exp\left(-\ln(2)\cdot\left(\frac{x-\mu_{x}}{\sigma_x}\right)^{p_x}\right), \nonumber
\end{equation}
 with $x=z,M$.
$\mu_x$, $\sigma_x$, and $p_x$ can be chosen for each template separately, while $p_x$ is even.
The individual photometric redshift value of a galaxy is determined by the mode of the $P(z)$ distribution $z_\mathrm{mode}$ of the best fitting model.
We estimate the error of the individual \gls{photoz} prediction to 
\begin{equation}
\label{eqn:paper2PhotoZerrors}
  \delta z=\left(\sum_z(z-z_{\mathrm{best}})^2\cdot P(z)\right)^{1/2} 
\end{equation}
Furthermore, we determine the stacked probability densities for all models in the template set which we will use in Sec.~\ref{sec:BOSSphotozresults} in the calculation of \gls{photoz} quality metrics.\\
The \cphotoz\ code has been applied in the past to a variety of photometric catalogues \citep{2001MNRAS.325..550D, 2004A&A...421...41G, 2005ApJ...619L.131D, 2005ApJ...633L...9F, 2008arXiv0811.3211B, 2008MNRAS.383.1319G, 2013ApJ...768..117G, 2013MNRAS.432.1046B, 2013MNRAS.432.1455G, 2014MNRAS.442.1507G, 2014MNRAS.445.1482S}.
It is also part of the PS1 Photometric Classification Server \citep{2012ApJ...746..128S}.


\section{The BOSS CMASS Sample}
\label{sec:paper2BOSSCMASSsample}
In this paper we use the spectroscopic data from the \emph{Baryon Oscillation Spectroscopic Survey} \citep[BOSS,][]{2011AJ....142...72E,2013AJ....145...10D}.
The latest public data release of \gls{BOSS} \citep[DR10,][]{2014ApJS..211...17A} targets 14,555 square degrees of the sky obtaining spectra for 1,848,851 galaxies.
The two spectrographs used in \gls{BOSS} are rebuilt from the original \gls{SDSS} spectrographs \citep{2013AJ....146...32S} and cover a wavelength range of 3600 to 10,400\,\Ang\ at a resolution of 1560 to 2650\,\Ang.
\gls{BOSS} was designed to measure the spatial distribution of \glspl{LRG} and quasars to investigate the baryon acoustic oscillations that are imprinted on the large scale structure of today's universe.
Galaxies surveyed by \gls{BOSS} have redshifts up to $z\sim1.0$.
The target selection is discussed in \citet{2011AJ....142...72E}.
\gls{BOSS} target galaxies are selected from \gls{SDSS}-II imaging data in such a way that they have high luminosities and masses.
Furthermore, the \gls{BOSS} target selection requires approximate uniform stellar masses throughout the redshift range of $0.2\lesssim z\lesssim0.6$.
Above $z\sim0.6$, the \gls{BOSS} sample is magnitude limited.
Similar to the cuts of the \gls{SDSS}-II \gls{LRG} sample \citep[][]{2001AJ....122.2267E,2005MNRAS.359..237P}, a number of magnitude and colour cuts are applied to ensure the above requirements are fulfilled.
They are selected in such a way to track the colours of a passively evolving galaxy from \citet[][M09 hereafter]{2009MNRAS.394L.107M}.
The \gls{BOSS} sample is divided into two subsamples, a lower redshift sample tagged \emph{LOWZ} at $z\leq0.4$, and a higher redshift sample, dubbed \emph{CMASS} for the constant mass requirement \citep{2011AJ....142...72E}.\\
The \gls{SDSS}-III data can be acquired from the \gls{SDSS} \ttt{CasJobs} website\footnote{http://skyserver.sdss3.org/CasJobs/}.
For the selection of our sample we employ the \gls{SDSS} \ttt{clean} photometry flag.
This flag ensures that we do not have duplicates in our sample by removing multiple detections on different frames.
Also, objects with deblending problems are removed, as well as the ones where more than 20\,\% of the PSF flux is interpolated over (that is, only 80\,\% of the flux is actually detected).
Additionally to that, we demand that objects were detected in the first pass (unbinned image), are not saturated, and that a radial profile could be constructed.\\
We require furthermore that galaxies in our sample have spectroscopic redshifts determined to a high accuracy.
Therefore, we chose only objects for our catalogue whose spec-$z$ warning flag is equal to zero.
We want to create models specifically designed for galaxies at higher redshifts than in G13.
Also, CMASS galaxies are very sparse at redshifts $z\gtrsim0.8$, which is why we reduce the sample to galaxies at $0.45\leq z\leq0.9$.
The resulting catalogue then contains $486,934$ CMASS galaxies.
The \ttt{CasJobs} SQL query used to acquire this catalogue is given in App.~\ref{app:paper2query}.\\
\begin{figure}
  \centering
  \includegraphics[width=\columnwidth]{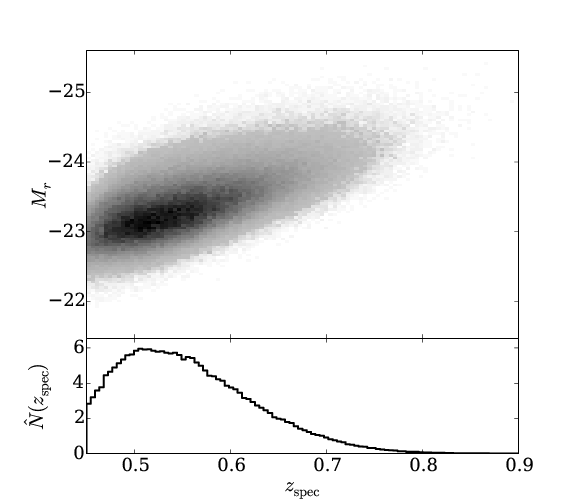}
  \caption{
    Upper panel: Spectroscopic redshift versus absolute magnitude in the $r$ band derived from a fit of G13 \gls{LRG} models to the CMASS catalogue of this work.
      Lower panel: Normalized frequency in spectroscopic redshift of the catalogue.
    }
  \label{fig:paper2zmabszspec}
\end{figure}
Fig.~\ref{fig:paper2zmabszspec} shows the redshift distribution of the such selected CMASS galaxies.
In the upper panel we plot the density of the sample in $\mzs$ versus absolute magnitude in the $r$ band.
$M_r$ is derived by fitting the \gls{LRG} models of G13 to the data at known spectroscopic redshifts.
The lower panel of Fig.~\ref{fig:paper2zmabszspec} presents the frequency in \sz, normalised to an integral of one.

\subsection{Colours of SED Templates versus Colours of BOSS Galaxies}
\begin{figure*}
  \centering
  \includegraphics[width=0.7\textwidth]{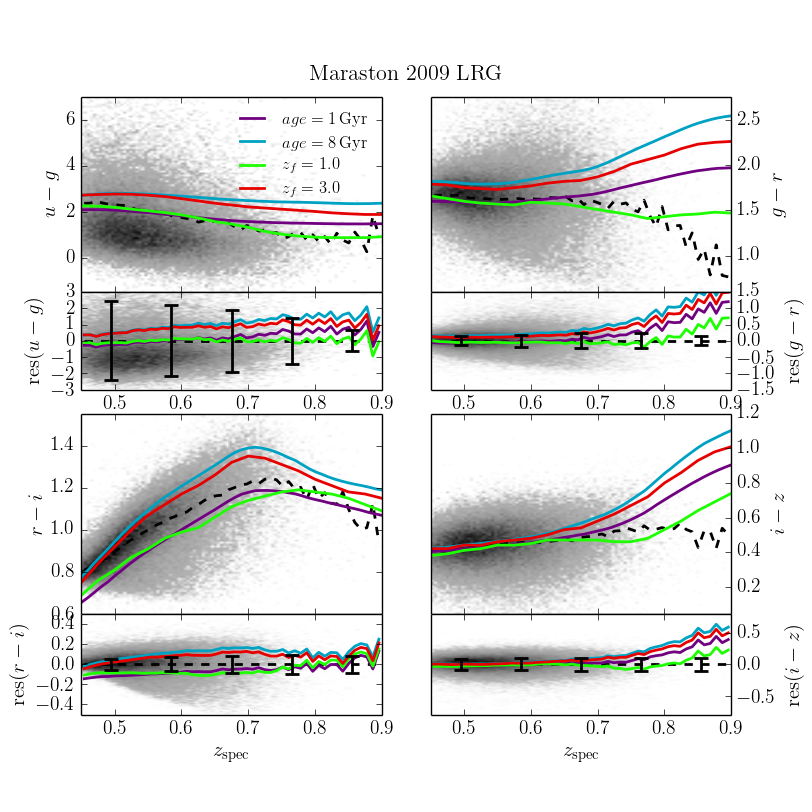}
  \caption{
    In the upper panels of the four major panels the CMASS colours are plotted as a function of redshift in grey, where the median is shown by a black dashed line.
    The lower panels are normalised to the median colour and show the residuals.
    Error bars show the median data errors in five redshift bins.
    On top of that the predicted colours of the M09 are plotted for several ages and formation redshifts.
    The purple and blue lines show the M09 model at constant ages of $1$ and $8\Gyr$.
    Furthermore, the green and red lines show a passively evolving version of M09 at formation redshifts of $z_f=1.0$ and $3.0$.
    }
  \label{fig:BOSSzcolM09}
\end{figure*}
In this section we want to compare the predicted colours of model \glspl{SED} to the colours of the \gls{BOSS} CMASS sample.
In Fig.~\ref{fig:BOSSzcolM09} we show the case of the M09 \gls{LRG} model which is used in the definition of the colour cuts of the \gls{BOSS} galaxy sample.
This model was created by M09 by adding a mass fraction of $3\,\%$ of metal poor stars ($\mZsun/200$) to a \gls{SSP} model with solar metallicity from the \citet{1998PASP..110..863P} stellar library in order to match the $g-r$ and $r-i$ colours of \gls{SDSS}-II \glspl{LRG}.
We sample the M09 model at ages of $1.0$ and $8.0\Gyr$ and predict the colours while redshifting the \gls{SED} (keeping the age fixed).
Additionally, we consider two passively evolving variations of the M09 model with formation redshifts of $z_f=1.0$ and $3.0$.
Since the M09 model is available only at distinct ages, we have to interpolate between them to determine the \glspl{SED} as a function of redshift.
The passive evolution was computed with the \cezgal\ software \citep{2012PASP..124..606M}.
M09 specifically created this model to match the colours of the \gls{SDSS}-II \glspl{LRG}, and it fits their median colours well (see also Fig.~5 in G13, or Figs.~1 and 2 in M09).
This is also true for the colours of the CMASS sample, where we find that the model with $z_f=1.0$ fits the \umg, \gmr, and \imz\ colours best in the observed cases.
In the case of the \gmr\ and \imz\ colours there are deviations from the data for $z\gtrsim0.8$, but also the sample size of the CMASS galaxies in this region decreases significantly.
However, the predicted colours in \rmi\ are too blue up to a redshift of $z\sim0.72$ and lie outside the median uncertainties of the data.
Furthermore, the colour errors are too small to be responsible for the spread in colour, wherefore the colour values of the data is due to the galaxy population not the photometric uncertainties.\\
Therefore, the data cannot be matched by the M09 model simultaneously in all colours, also not when taking evolution effects into account.
The spread in colour can be explained by differences in ages of the galaxy, but could be also due to different stellar populations, i.e., the metallicity and the distribution of stellar ages.

\section{New SED Templates}
\label{sec:paper2newsedtemplates}
We aim to create a set of templates that can be used for photometric redshift estimation of galaxies with similar properties as the CMASS sample.
To create models specifically designed to match the colours of the \gls{BOSS} data, we fit a number of model \glspl{SED} to the data at their known spectroscopic redshift and select from the best fitting models a subset that should represent the data in terms of colours, while yielding accurate photometric redshifts.\\
We expect the galaxy population to vary as a function of redshift.
In the process of \gls{photoz} estimation with template fitting we can account for that by assigning different redshift prior probabilities to individual model \glspl{SED}.
One could be tempted to use a huge amount of \glspl{SED} with different properties that can match the data in question and yield reliable \gls{photoz} results (while heavily increasing the computation time).
However, this is not the case since one has to deal with degeneracies in colours resulting from different galaxy properties, e.g., age and metallicity.
Also, introducing peculiar templates can deteriorate the $P(z)$ of a galaxy.
Therefore, one has to carefully select a small set of templates able to match the galaxy catalogue in question.
In order to create a set of models that match the data at different redshifts we fit a variety of model \glspl{SED} to the CMASS galaxies within four redshift bins.
The $z$ bins are centred on $z=0.5,0.6,0.7,0.8$ (continuing the sequence from G13) with interval widths of $\Delta z=0.04$.\\
In the following we first give a short introduction of the \gls{SED} fitting procedure and turn afterwards to the description of the models we used as a basis to construct new model \glspl{SED}.
After that we explain how we select models to represent the data that should then serve as a template set for photometric redshift estimation.

\subsection{Generating Model SEDs for Objects in the BOSS Catalogue by SED Fitting}
\label{sec:paper2SEDfitting}
To fit model \glspl{SED} to the data we use the \gls{SED} fitting software \csedfit\ \citep{2004ApJ...616L.103D}.
The code fits a number of model \glspl{SED} to the colours of the data by performing a maximum likelihood fit.
Additionally to the models, one can define a burst model which is then fitted in superposition to the main component at several mass fractions.
Furthermore, the \csedfit\ code applies dust extinction to both, the main and star burst component, following the extinction law of \citet{2000ApJ...533..682C}.
We perform the \gls{SED} fitting procedure in such a way that signal-to-noise ratios smaller than $1$ are considered upper limits.\\ 
The \br\ is the most significant feature in the spectra of red galaxies and the reason why we can estimate \glspl{photoz} from broad band photometry to a high accuracy.
At redshifts populated by the CMASS galaxies the \br\ lies within the $r$ band, wherefore the $g$ band is the bluest band needed to derive the position of the break in wavelength.
The $u$ band is very shallow and it may deteriorate the quality of the \gls{SED} fitting results and the photometric redshifts if the errors are not determined accurately.
For these reasons we decide to omit it in the following \gls{SED} fitting and photometric redshift estimation.
Furthermore, the \gls{SED} fits are not performed on the whole data of a $z$ bin because of the high computation time, but we randomly select subsamples from within each $z$ bin containing ${\sim}\,2000$ objects.

\subsubsection{SED Fitting Results with BC03 Models}
\label{sec:paper2SEDfittingoriginalBC03}
We tried a variety of available model \glspl{SED} in the \gls{SED} fitting procedure.
The most extensive public libraries originate from \citet[][hereafter dubbed BC03]{2003MNRAS.344.1000B}, \citet[][hereafter M05]{1998MNRAS.300..872M,2005MNRAS.362..799M}, and \citet[][hereafter M11]{2011MNRAS.418.2785M}.
We fit all of these models in separate trial runs in order to select a basic model set to proceed further, since large sets of models require enormous computation times.\\
The BC03 models can be generated by the software \cgalaxev\footnote{http://www.cida.ve/\textasciitilde bruzual/bc2003 or http://www.iap.fr/\textasciitilde charlot/bc2003}.
We create synthetical models from the BaSeL\,3.1 library, using the Padova\,1994 evolutionary tracks recommended by BC03 and the \gls{IMF} by \citet{2003PASP..115..763C}.
We generate models with four different metallicities, $Z=0.2$, $0.4$, $1$, and $2.5\,\mZsun$.
From these we produce \gls{SSP} models, as well as \gls{CSP} models.
\glspl{SSP} assume that all stars are formed instantaneously in a delta-functional star burst at the birth of the galaxy (age zero) and are passively evolving afterwards.
To create \glspl{CSP} one can assign essentially any function for the star formation history (SFH), such that star formation takes place for longer periods of time.
Usually, a star formation rate is assumed that behaves proportional to $\exp(-t/\tau)$ (at least for galaxies at lower redshifts $z\lesssim1$), where $t$ is the age of the galaxy and $\tau$ is the $e$-folding time scale \citep[e.g.,][]{2005ApJ...626..698S,2009MNRAS.394..774L}.
We produce models with different $\tau$ values, $\tau=0.5, 1.0, 3.0, 5.0, 8.0, 20, 50\,\Gyr$, where the latter simulate an almost constant \gls{SFR}.
We sample the models at ages ranging from $10\,\rm{Myr}$ to $13\Gyr$.
In order to exploit the maximum freedom available, we also create model \glspl{SED} with increasing \gls{SFR}, hence $\tau<0$.
This kind of \gls{SFH} is mostly important for high(er) redshifts ($z\gtrsim2$) and yields more physical results \citep[e.g.,][]{2010MNRAS.407..830M,2014MNRAS.438.1417M}, but was used for \gls{SED} fitting also at lower redshifts \citep[e.g.,][]{2012MNRAS.422.3285P}.
In our case the increasing \gls{SFR} models do not significantly increase the range in colour space needed to match the CMASS galaxies which is why we omit them in the following to save computation time.\\
We create the same variety of \glspl{CSP} for the M05 and M11 models (at least where possible) using the stellar population synthesis code \cezgal\ \citep{2012PASP..124..606M}, and fit them to the data as well.
We show the distribution in likelihoods in Fig.~\ref{fig:chi2distr} in App.~\ref{app:likelihood} for the \gls{SED} fits of BC03, M05, and M11 models to the CMASS data in the four redshift bins.
The BC03 models outperform the M05 and M11 models in terms of their $\chi^2$ values marginally at most.
Fig.~\ref{fig:chi2distr} shows that the fits with BC03 models have a slightly higher frequency of lower $\chi^2$ values than M05/11.
Furthermore, considering Fig.~20 in G13, the BC03 models are a better match to the \gls{SDSS}-II \gls{LRG} data in terms of $\chi^2$ values (this is not so evident here, as seen in Fig.~\ref{fig:chi2distr}).
Because of the versatility in the model creation with \cgalaxev, and the variety of provided ages and metallicities, we choose to use the BC03 models in the remainder of this work.\\
\begin{figure}
  \centering
  \includegraphics[width=\columnwidth]{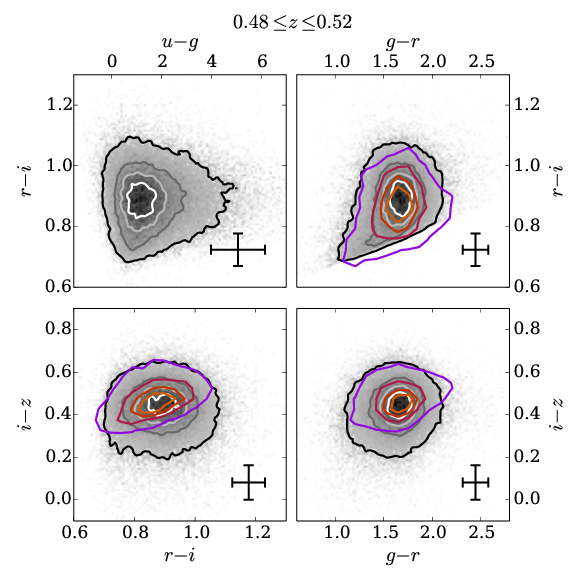}
  \caption{
    Colour-colour plots for CMASS galaxies at $0.48\leq z\leq0.52$.
    The \gls{BOSS} data is shown by grey shades and accompanying grey contours.
    Error bars denote the median error in colour in this redshift bin.
    The density of the resulting colours of the \gls{SED} fitted unmodified BC03 models are shown by coloured contours.
    Both sets of contours are drawn at the same frequency levels, $5\,\%$, $25\,\%$, $45\,\%$, and $65\,\%$.
  }
  \label{fig:paper2SEDfitcolcolbc03orig5}
\end{figure}

\begin{figure}
  \centering
  \includegraphics[width=\columnwidth]{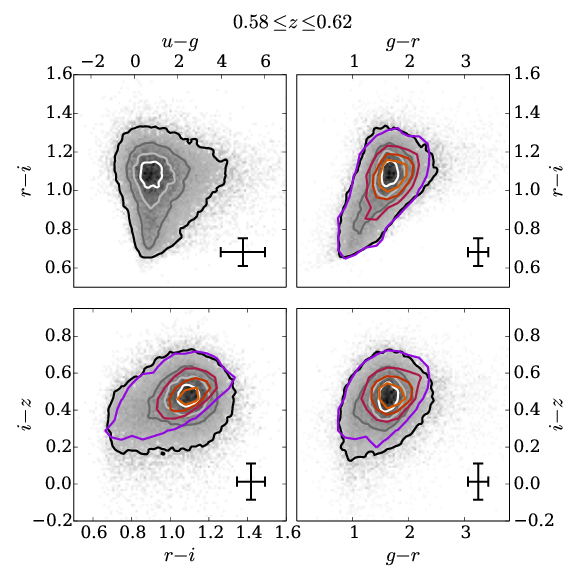}
  \caption{
    Colour-colour plots for CMASS galaxies and best fitting BC03 models at $0.58\leq z\leq0.62$.
    See Fig.~\ref{fig:paper2SEDfitcolcolbc03orig5} for a detailed description.
  }
  \label{fig:paper2SEDfitcolcolbc03orig6}
\end{figure}

\begin{figure}
  \centering
  \includegraphics[width=\columnwidth]{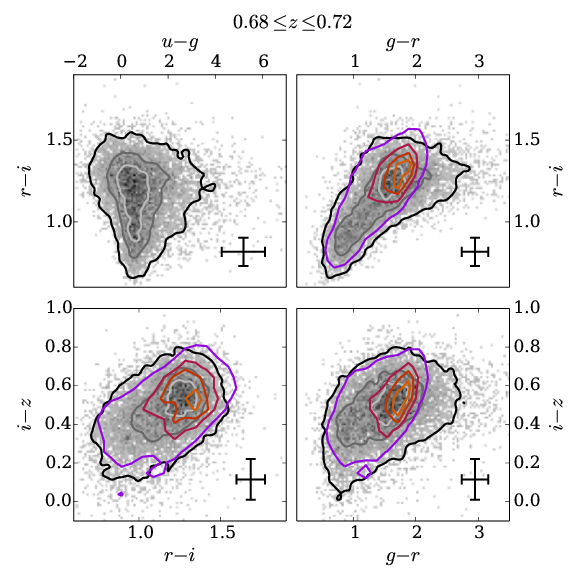}
  \caption{
    Colour-colour plots for CMASS galaxies and best fitting BC03 models at $0.68\leq z\leq0.72$.
    See Fig.~\ref{fig:paper2SEDfitcolcolbc03orig5} for a detailed description.
  }
  \label{fig:paper2SEDfitcolcolbc03orig7}
\end{figure}

\begin{figure}
  \centering
  \includegraphics[width=\columnwidth]{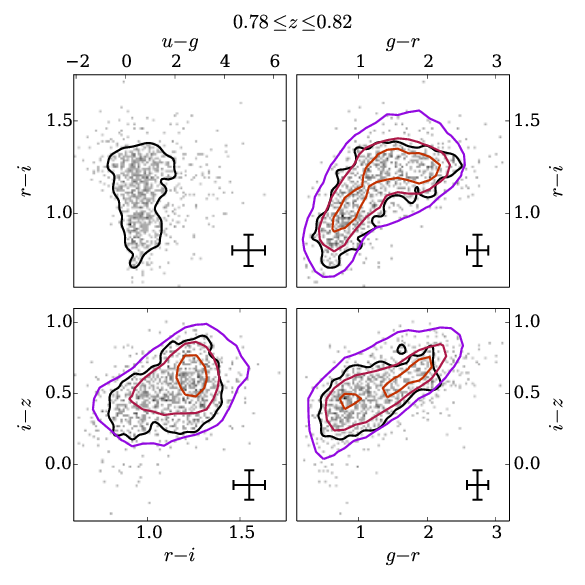}
  \caption{
    Colour-colour plots for CMASS galaxies and best fitting BC03 models at $0.78\leq z\leq0.82$.
    See Fig.~\ref{fig:paper2SEDfitcolcolbc03orig5} for a detailed description.
    The contours had to be smoothed to make them visible due to the small sample size in this redshift bin.
  }
  \label{fig:paper2SEDfitcolcolbc03orig8}
\end{figure}
  
In Figs.~\ref{fig:paper2SEDfitcolcolbc03orig5} to \ref{fig:paper2SEDfitcolcolbc03orig8} we present the \gls{SED} fitting results with BC03 \gls{SSP} and \gls{CSP} models with increasing and decreasing \glspl{SFR} in colour space.
The data colour is plotted in grey scales with according contours, and the median colour errors are shown by error bars in each panel.
The density of the predicted colours of the best fitting models are plotted over the data distribution in coloured contours.
Both sets of contours are plotted at the same frequency levels.\\
The bluer \imz\ colour range populated by the CMASS galaxies in the redshift bins at $z\sim0.5$. $z\sim0.6$ and $z\sim0.7$ is not populated by the best fitting BC03 models.
Additionally, the redder \gmr\ colours at $z\sim0.7$ are also not matched by the BC03 models.
All these offsets in colour cannot be accounted for by the photometric errors which are also given in Figs.~\ref{fig:paper2SEDfitcolcolbc03orig5} to \ref{fig:paper2SEDfitcolcolbc03orig8}.
We investigate the colour mismatch in the next section and modify the model \glspl{SED} to fit the data to a better extent.

\subsubsection{Modification of the Red Continuum Slope}

In G13 we already showed in Fig.~19 that the predicted \imz\ colour of the \gls{SDSS}-II \gls{LRG} sample cannot be matched by the models for the highest redshift bin ${\sim}\,0.4$, at least not while the other three colours are fitted simultaneously.
The model colours were too red in comparison to the data, which means that the decrease in the continuum slope redwards of the \br\ of the model \glspl{SED} is not strong enough.
The mismatch in \imz\ is also present for the CMASS sample which is located at even higher redshifts than the \gls{SDSS}-II \glspl{LRG}.\\
The slope of the continuum is changed by variations in the physical properties of the models which we investigate in the following paragraph.
It is unlikely that the choice of the \gls{IMF} could affect the model \glspl{SED} in such a way to produce the colour mismatch we observe.
Changing the Chabrier \gls{IMF} to a \citet{1955ApJ...121..161S} or \citet{2001MNRAS.322..231K} \gls{IMF} would only result in a change of the abundance of low mass stars.
These should not have a great impact on the continuum slope, since the light in this part of the spectrum is dominated by red giants.
Optical colours are not sensitive to the choice of the IMF, which is (when stars are not resolved) often derived from spectroscopy or IR photometry \citep[e.g.,][]{2003MNRAS.339L..12C,2012ApJ...744...17M,2012ApJ...760...71C}.
In turn, the choice of the particular IMF cannot cause the observed mismatch in colours.
Different metallicities change the slope in the continuum as well, but we have considered sub- to super-solar values.
Furthermore, we have exploited the model age and extinction as further parameters up to still reasonable values, i.e., $13\,\Gyr$ and $\mathbf{A_V}=3.0$.
The burst component significantly affects only the bluer parts of the spectrum and has only marginal impact on the red part, which is dominated by the main stellar population, and not by the small (in total mass) burst fractions.
Lastly, the \gls{SFH} has to be considered.
We analysed \glspl{CSP} with nearly constant \gls{SFR} and \glspl{SSP} which create all their stars at one instant in time.
We even considered exponentially increasing \glspl{SFR}.
If the too shallow red continuum slope would be a result of a poorly chosen \gls{SFH}, we would at least expect that the data is bracketed by the considered models in colour space.\\
Since we cannot isolate physical model parameters that cause the colour mismatch, we can only assume that the issue arises due to inaccuracies in the modelling of the stellar evolution phase most difficult to follow theoretically, i.e., the \gls{AGB}.
The wavelength ranges covered by the colour mismatch hints at an incorrect modelling of \gls{AGB} stars, since these dominate the SED in the red optical and NIR parts of the spectrum.
E.g., \citet{2013MNRAS.428.1479Z} show through NIR spectroscopy (comparing BC03 and M05 models in their Fig.~1) that the chemical composition influences the slope of the model SEDs in the NIR, i.e., that these differences are caused by different C/O abundances.
Furthermore, the different composition of dust in the circumstellar envelopes of carbon and oxygen-rich \gls{AGB} stars affect the efficiency of dust absorption and emission which can therefore greatly influence the colours \citep[e.g.,][]{2008A&A...482..883M,2013MNRAS.436.2824C,2014A&A...565A...9S}.
Other explanations could be that the simple mixing-length theory models describing the convection in stars in the \gls{AGB} phase are insufficient to reproduce the colours, or that incorrect assumptions on mass loss leads to the deviation.\\
A close investigation of the modelling of the \gls{AGB} phase is needed to resolve the problem which is beyond the scope of this paper.
However, we explain in the next paragraph how we modify the red continuum slope to better match the data.\\
The red \gls{SED} continuum follows a function proportional to $\lambda^{-\beta^\prime}$, heavily modified by absorption lines.
Therefore, we can also change the steepness of the continuum by varying $\beta^\prime$.
This is done in such a way that we multiply the \gls{SED} redwards of a wavelength $\lambda_i$ with $a\cdot\lambda^{-\beta}$, where $a$ is chosen such that the fluxes of the underlying model \gls{SED} and the modified \gls{SED} coincide at $\lambda_i$.
In this way we can change the continuum slope for $\lambda\geq\lambda_i$, wherefore the created \gls{SED} on average follows $\lambda^{-(\beta+\beta^\prime)}$.
We will term the such modified models ``$\lambda^{-\beta}$ models'' in the remainder of the text.\\
\begin{figure}
  \includegraphics[width=\columnwidth]{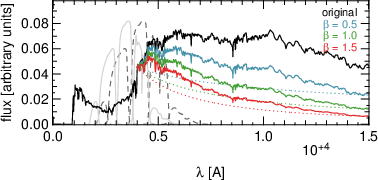}
  \caption{
    The black line is an \gls{LRG} model \gls{SED} from G13.
    The dotted blue, green, and red lines show the function $\lambda^{-\beta}$ with $\beta=0.5,1.0,1.5$ for $\lambda\geq4000\mAng$.
    The solid lines are the above \gls{SED} multiplied by $\lambda^{-\beta}$, also for $\lambda\geq4000\mAng$.
    In grey (dark grey dashed line, and light grey solid line) the \gls{SDSS} $ugriz$ filter curves are shown for a galaxy at $z=0.5$, and $z=0.8$ respectively.
  }
  \label{fig:expbeta}
\end{figure}
We exemplary show the resulting \glspl{SED} of an \gls{LRG} model from G13 when modified by $a\cdot\lambda^{-\beta}$ for $\lambda\geq\lambda_i=4000\,\mAng$ and $\beta=0.5,1.0,1.5$ in Fig.~\ref{fig:expbeta}.
We also plot the \gls{SDSS} filters as positioned in wavelength in the observed frame at $z=0.5$ and $z=0.9$, the approximate extreme redshift values of the CMASS sample.
We can see that for $\lambda\geq4000\,\mAng$ the change in slope mostly affects the $i$ and $z$ band at lower redshifts.
For higher redshifts $\lambda_i=4000\,\mAng$ has moved in the $r$ band.
Therefore, we create variations of the BC03 \gls{SSP} and the \gls{CSP} models with decreasing \gls{SFR} explained above, with $\lambda_i\in[3000,3500,4000,4500,5000,5500,6000]\,\mAng$ such that the variation does affect the fluxes in different filters for similar redshifts.
Furthermore, the $\beta$ values span a range from $0.5$ to $2.0$ with a step size of $0.5$.
Larger ranges for values of $\lambda_i$ and/or $\beta$ do not improve the results further, since the location in colour spaces they would inhabit is already covered by the other models or not populated by the CMASS galaxies.\\
\begin{figure*}
  \includegraphics[width=0.7\textwidth]{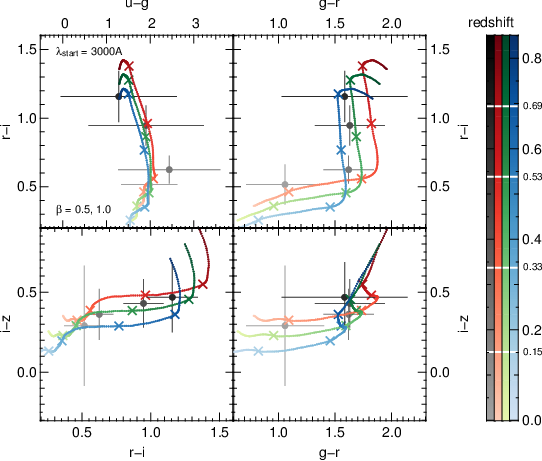}
  \caption{
    Colour-colour plots of the CMASS data and model \glspl{SED}.
    The \gls{BOSS} data is split into four equally sized redshift bins.
    From these bins we calculate the mean colour and plot it as points in the panel.
    The error bars denote the accompanying root mean square values of the colour in each bin.
    The points and the error bars are colour-coded with redshift (see the colour bar on the right, where white lines indicate the mean redshift in each bin).
    The red shaded track shows the colours of the model \gls{SED} of Fig.~\ref{fig:expbeta} with varying redshift, where the crosses are at the same $z$ as the data.
    Again, the redshift of the track is colour coded and shown in the colour bar.
    The green and blue shaded tracks originate from the same model, when multiplied by \expb\ with $\beta=0.5,1$ for $\lambda\geq3000\,\mAng$.
    A more detailed description is given in the text.
  }
  \label{fig:expbetacol}
\end{figure*}
In Fig.~\ref{fig:expbetacol} we show the colours of the CMASS data (grey).
We furthermore plot the colours of an exemplary \gls{SED} from G13 (red), and the colours of the same model when modified by $\lambda^{-\beta}$, for $\beta=0.5$ and $1$ (green and blue).
The data in Fig.~\ref{fig:expbetacol} is split into four redshift bins, and we plot the mean colour and corresponding root mean square values of each bin.
The points are colour-coded in such a way that they darken with increasing redshift, which is shown by the grey colour bar on the right of Fig.~\ref{fig:expbetacol}, where white lines denote the mean redshifts within the bins.
On top of the data, the colour tracks as functions of redshift are plotted for the models.
The model colours at the $z$ values of the data are highlighted by crosses, and the tracks in the colour spaces are again colour-coded as a function of redshift.
The track of the original model is plotted by varying intensities of red, whereas those of the corresponding \expb\ models are plotted in green shades for $\beta=0.5$, and blue shades for $\beta=1$ respectively.
All plotted models have $\mli=3000\,\mAng$.
While the data colours (grey) are well matched by the original model (red track) at lower redshifts, the deviations from the mean colour increase with increasing $z$.
The modification by \expb\ produces a better agreement with the mean data colours.
The value of $\beta$ necessary to fit the data best is somewhat ambiguous in this plot, but has to be decided individually for each galaxy together with the best fitting model.


\subsubsection{SED Fitting Results with modified BC03 Models}
\label{sec:paper2SEDfittingresults}

\begin{figure}
  \includegraphics[width=\columnwidth]{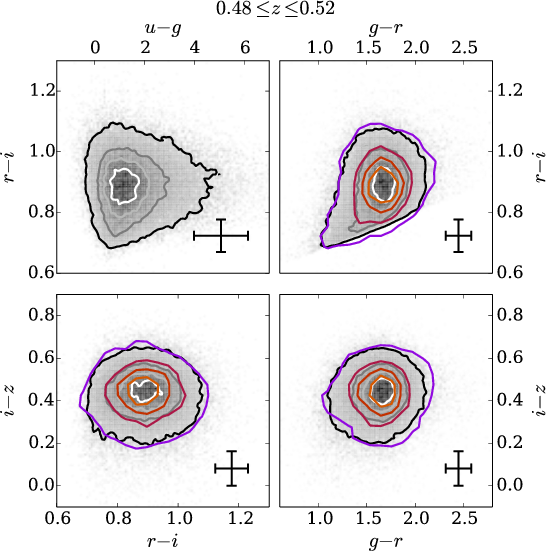}
  \caption{
    Colour-colour plots for CMASS galaxies at $0.48\leq z\leq0.52$ (analogous to Fig.~\ref{fig:paper2SEDfitcolcolbc03orig8}).
    The \gls{BOSS} data is shown by grey shades and accompanying grey contours.
    Error bars denote the median error in colour in this redshift bin.
    The density of the resulting colours of the best fitting BC03 and \bcexpb\ models are shown by coloured contours.
    Both sets of contours are drawn at the same frequency levels, $5\,\%$, $25\,\%$, $45\,\%$, and $65\,\%$.
    }
  \label{fig:paper2SEDfitcolcolbc03expbeta5}
\end{figure}

\begin{figure}
  \includegraphics[width=\columnwidth]{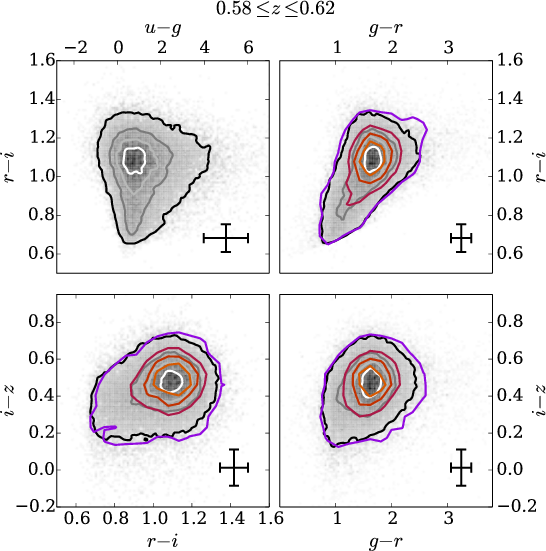}
  \caption{
    Colour-colour plots for CMASS galaxies and best fitting \bcexpb\ models at $0.58\leq z\leq0.62$ (analogous to Fig.~\ref{fig:paper2SEDfitcolcolbc03orig6}).
    See Fig.~\ref{fig:paper2SEDfitcolcolbc03expbeta5} for a detailed explanation of the plot.
  }
  \label{fig:paper2SEDfitcolcolbc03expbeta6}
\end{figure}

\begin{figure}
  \includegraphics[width=\columnwidth]{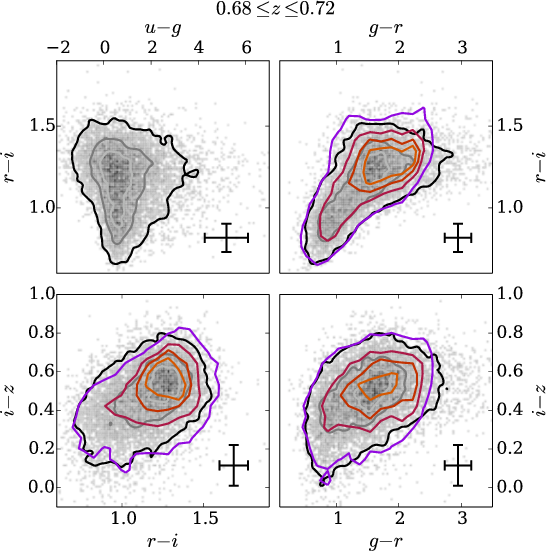}
  \caption{
    Colour-colour plots for CMASS galaxies and best fitting \bcexpb\ models at $0.68\leq z\leq0.72$ (analogous to Fig.~\ref{fig:paper2SEDfitcolcolbc03orig7}).
    See Fig.~\ref{fig:paper2SEDfitcolcolbc03expbeta5} for a detailed explanation of the plot.
  }
  \label{fig:paper2SEDfitcolcolbc03expbeta7}
\end{figure}

\begin{figure}
  \includegraphics[width=\columnwidth]{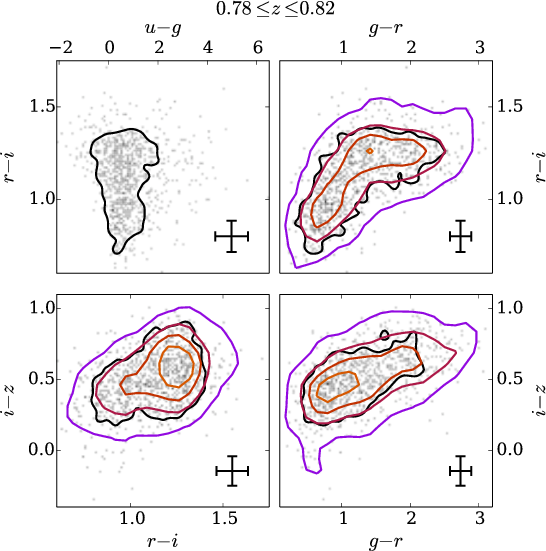}
  \caption{
    Colour-colour plots for CMASS galaxies and best fitting \bcexpb\ models at $0.78\leq z\leq0.82$ (analogous to Fig.~\ref{fig:paper2SEDfitcolcolbc03orig8}).
    See Fig.~\ref{fig:paper2SEDfitcolcolbc03expbeta5} for a detailed explanation of the plot.
    The contours had to be smoothed to make them visible due to the small sample size in this redshift bin.
    This is done in the same way as in Fig.~\ref{fig:paper2SEDfitcolcolbc03orig8}.
  }
  \label{fig:paper2SEDfitcolcolbc03expbeta8}
\end{figure}

We now introduce the modified versions of the BC03 models together with the original \glspl{SSP} and \glspl{CSP} with decreasing \glspl{SFH} from Sec.~\ref{sec:paper2SEDfittingoriginalBC03} in the \csedfit\ code with the same fitting parameters as in Sec.~\ref{sec:paper2SEDfittingoriginalBC03}.
Figs.~\ref{fig:paper2SEDfitcolcolbc03expbeta5} through \ref{fig:paper2SEDfitcolcolbc03expbeta8} show the \gls{SED} fitting results in colour spaces of the CMASS data in the four $z$ bins.
We see that the contours of the data and the best fitting models (which are located at the same steps in frequency) almost coincide in Figs.~\ref{fig:paper2SEDfitcolcolbc03expbeta5}, \ref{fig:paper2SEDfitcolcolbc03expbeta6}, and \ref{fig:paper2SEDfitcolcolbc03expbeta7}.\\

\begin{figure*}
  \centering
  \includegraphics[width=0.7\textwidth]{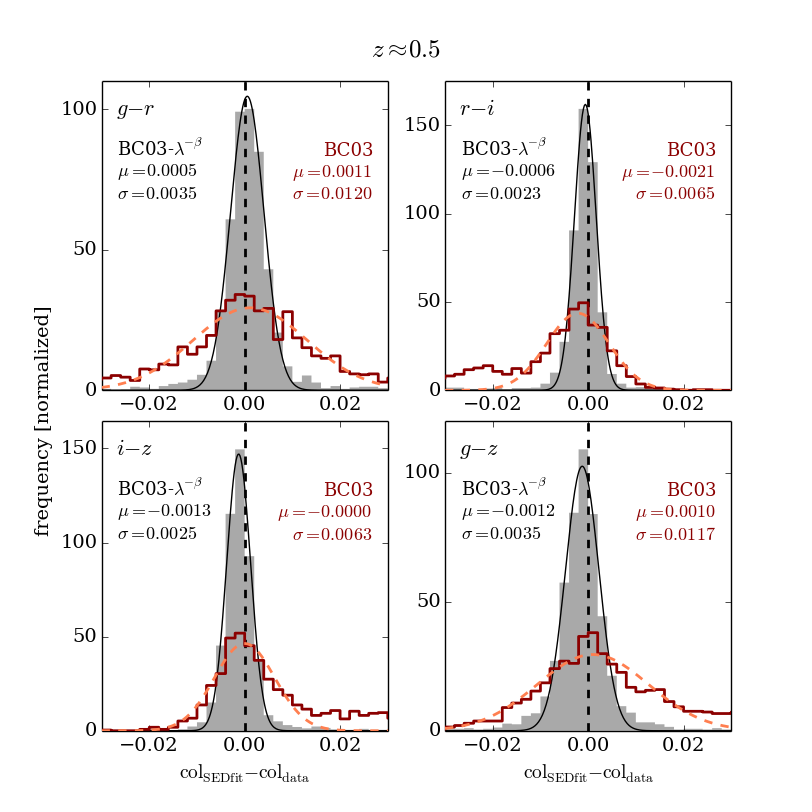}
  \caption{
    Deviations in \gmr, \rmi, \imz\ and $g-z$ colours for BC03 models and \bcexpb\  models from the data within $0.48\leq z\leq0.52$.
    The $\Delta\mathrm{col}$ distribution for the \bcexpb\  models are plotted in grey, and the best fitting Gaussian is indicated by a solid black line.
    The same distribution, but for unmodified BC03 models is given by a dark red line, where we plot the best fitting Gaussian by a dashed orange line.
    The black dashed vertical line highlights $\Delta\mathrm{col}=0$.
    The best fitting parameters of the Gaussian curves are given in the plots.
  }
  \label{fig:dcol5}
\end{figure*}

\begin{figure*}
  \centering
  \includegraphics[width=0.7\textwidth]{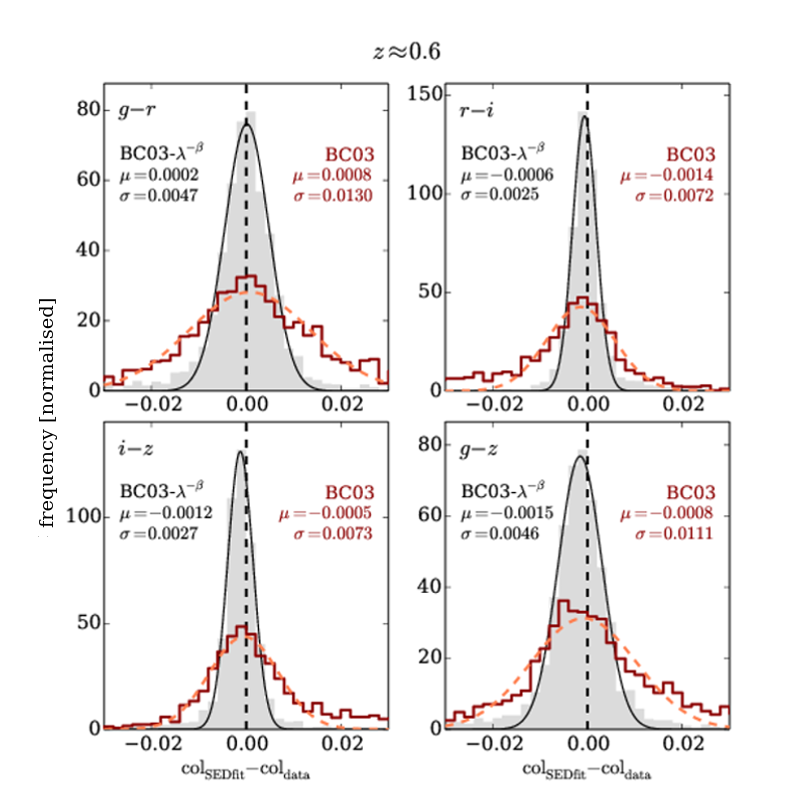}
  \caption{
    Deviations in \gmr, \rmi, \imz\ and $g-z$ colours for BC03 models and \bcexpb\  models from the data within $0.58\leq z\leq0.62$.
    A detailed description of the plot is given in Fig.~\ref{fig:dcol5}.
  }
  \label{fig:dcol6}
\end{figure*}

\begin{figure*}
  \centering
  \includegraphics[width=0.7\textwidth]{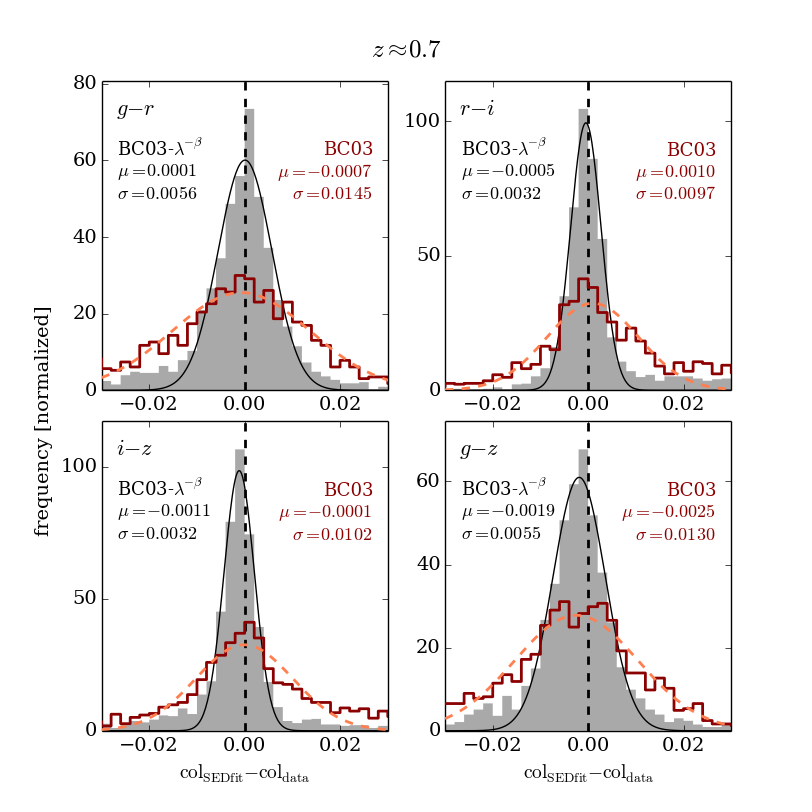}
  \caption{
    Deviations in \gmr, \rmi, \imz\ and $g-z$ colours for BC03 models and \bcexpb\  models from the data within $0.68\leq z\leq0.72$.
    A detailed description of the plot is given in Fig.~\ref{fig:dcol5}.
  }
  \label{fig:dcol7}
\end{figure*}

\begin{figure*}
  \centering
  \includegraphics[width=0.7\textwidth]{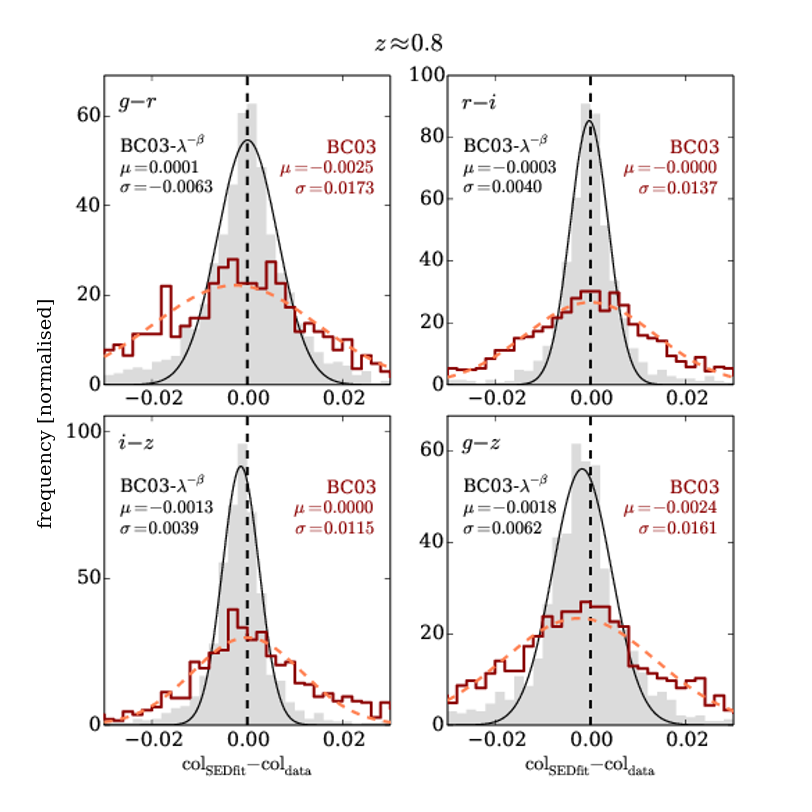}
  \caption{
    Deviations in \gmr, \rmi, \imz\ and $g-z$ colours for BC03 models and \bcexpb\  models from the data within $0.78\leq z\leq0.82$.
    A detailed description of the plot is given in Fig.~\ref{fig:dcol5}.
  }
  \label{fig:dcol8}
\end{figure*}

Comparing Figs.~\ref{fig:paper2SEDfitcolcolbc03orig5} to \ref{fig:paper2SEDfitcolcolbc03orig8} with Figs.~\ref{fig:paper2SEDfitcolcolbc03expbeta5} to \ref{fig:paper2SEDfitcolcolbc03expbeta8}, we can observe an improvement on the match between the locations in colour space populated by the best fitting models and the data.
This is additionally confirmed by Fig.~\ref{fig:chi2distr} in App.~\ref{app:likelihood}, where we see that the resulting $\chi^2$ values of the fits are more frequently found at lower $\chi^2$ values for the \bcexpb\ models than for the original ones, hence improving the goodness of fit.
Here, we want to specifically analyse the offsets in the \gmr, \rmi, \imz, and $g-z$ colours of the best fitting BC03 models and their modified variations from the data.
We show in Figs.~\ref{fig:dcol5} to \ref{fig:dcol8} the deviations of the model colours, predicted by the best fitting \glspl{SED}, to the data, $\Delta\mathrm{col}=\mathrm{col_{model}-col_{data}}$, for both setups and in all four considered $z$ bins.
The distribution in $\Delta\mathrm{col}$ of the original BC03 models (\glspl{SSP}, as well as decreasing and increasing \gls{SFR} \glspl{CSP}) is presented by red histograms, where the Gaussian curve best fitting the histogram is shown by a dashed orange line.
The colour offsets of the \bcexpb\ models are plotted by a grey histogram, with an according Gaussian drawn in black.
$\Delta\mathrm{col}=0$ is shown by a dashed black line, and we print the parameter values of the Gaussians in the plots.\\
While the mean deviations presented in Figs.~\ref{fig:dcol5} to \ref{fig:dcol8} are more or less the same in all considered redshift bins, the standard deviation of $\Delta\mathrm{col}$ is about $2$ to $3$ times higher for the original BC03 \glspl{SED}.
Furthermore, the \bcexpb\ distributions of $\Delta\mathrm{col}$ resemble a Gaussian much more closely than that of the original BC03, especially in the \rmi\ and \imz\ colours, but also in $g-z$.
We already pointed out that the \imz\ colour is overestimated by the BC03 models analysed in G13 for $z\sim0.4$.
This is still true for $z\sim0.5$ to $0.7$, observable in Figs.~\ref{fig:dcol5} to \ref{fig:dcol7} (cf. Figs.~\ref{fig:paper2SEDfitcolcolbc03expbeta5} to \ref{fig:paper2SEDfitcolcolbc03expbeta7}).
Concurrently, the predicted \rmi\ colours of the unmodified BC03 models are too blue up to $z\sim0.6$, and too red in $z\sim0.7$.
In contrast to that, the colours of the \bcexpb\ \glspl{SED} yield very symmetric distributions, although with increased flanks in comparison to a Gaussian for higher redshifts.\\

In summary, we investigated two additional degrees of freedom, apart from the model \gls{SED} parameters metallicity, \gls{SFH}, age, burst, and dust, by modification of the red continuum slope through multiplying with $\lambda^{-\beta}$ at $\lambda\geq\lambda_i$ to match the colours of the CMASS galaxies to a better extent.
When fitted to the data, the \expb\ \glspl{SED} predict colours that deviate less from the data than the unmodified models they originate from, and yield lower $\chi^2$ values.
We will use the variety of best fitting \glspl{SED} (including the additional burst and dust components) as a basis to select from when we define a model set for photometric redshift template fitting codes.

\subsection{Selection of Best Fitting SEDs for the New Template Set}
\label{sec:BOSSselection}
\begin{figure*}
  \centering
  \includegraphics[width=0.7\textwidth]{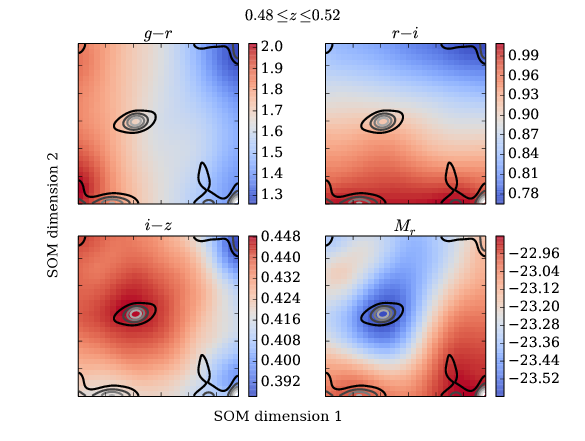}
  \caption{
    \gls{SOM} with $30\times30$ bins for \gmr, \rmi, \imz, and $M_r$ for galaxies within $0.48\leq z\leq0.52$, where the parameter values are colour-encoded (see colourbars on the right of each panel).
    $M_r$ is computed by fitting templates from G13 to the spectroscopic redshift.
    Contours show the number density of the data in the map (note that the contours are smoothed to improve the clarity of the figure) and are drawn at six equally distributed frequency levels from ${\sim}\,15\,\%$ (black) to ${\sim}\,85\,\%$ (white).
  }
  \label{fig:SOM5}
\end{figure*}
We want to select a set of model \glspl{SED} from the best fitting models of the previous section that represent the data in terms of colours for each redshift bin.
The space we can construct from the \gls{SDSS} colours is many-dimensional, and we want to reduce the dimensions for simplicity of the selection process.
Therefore, we decided that the target space should have two dimensions, a compromise between the loss of information (which is greater for less dimensions) and the simplification through the reduction of dimensions.
We could, in principle, perform a \gls{PCA}, and reduce the dimensions by concentrating on the space which is spanned by the first two eigenvectors of the \gls{PCA} that have the highest variance.
Another possibility for the reduction of dimensions is a self-organising map \citep[SOM or Kohonen-map,][]{SOM,2001som..book.....K}.
A \gls{SOM} is an artificial neural network (ANN) that provides a discrete representation of a set of higher dimensional data values in a lower dimensional space (most often two-dimensional).
The network is trained using unsupervised learning to map the data onto the surface in such a way that data points with similar properties (i.e., data values) are located in close neighbourhoods.
Unlike a \gls{PCA}, where the neglection of the third and later components yields a complete loss of the information carried by them, the two-dimensional \gls{SOM} still entails this information in its points.
This is why we chose a \gls{SOM} over a \gls{PCA} for the mapping onto a two-dimensional plane in the model selection below.\\  
We create maps of the galaxy catalogues from the \gmr, \rmi, and \imz\ colours of the four considered redshift bins.
Furthermore, we can easily introduce also the absolute magnitudes $M$ of the galaxies as a forth quantity to be mapped, since $M$ can also hold information about the colour \citep[e.g.,][]{2004ApJ...600..681B}.
We have to take care of the errors in the data by normalising the colours to their mean value and dividing through the colour errors determined by the uncertainties in the photometry.
The absolute magnitudes of the galaxies are calculated by fitting the \gls{LRG} model \glspl{SED} of G13 to the data at their spectroscopic redshifts.
From the best fitting models we extract the absolute magnitude in the \gls{SDSS} $r$ band, which has to be normalised to a standard normal distribution to be comparable to the colour values.
Once this is done we train the \gls{SOM} and create a surface with $30\times30$ discrete $x$- and $y$-values with assigned input values (i.e., colours and $M_r$).
The positions of data points in the \gls{SOM} are then determined by performing a nearest neighbour search.\\
We use a \gls{SOM} algorithm implemented in python and provided in the \ttt{PyMVPA} package \citep{PyMVPA}.
Fig.~\ref{fig:SOM5} presents the \gls{SOM} trained on the catalogue with $0.48\leq z\leq0.52$ on the \gmr, \rmi, and \imz\ colours and on $M_r$.
The four panels are representations of the same \gls{SOM}, but with the values of the four mapped parameters \gmr, \rmi, and \imz\ colours and $M_r$ encoded in colour.
The density of the mapped underlying data are shown by contours.
We can see from Fig.~\ref{fig:SOM5} that the extreme values of \gmr\ and \rmi\ have the greatest separation in the map, which is due to these colours having the largest spread in values.
This is not a result of measurement errors, since we took these into account through the normalisation of the colours explained in the last paragraph.
We can see for example that the dense region in the middle of the panels corresponds to a higher concentration of the data at the respective colour values.
These are $g-r\sim1.7$, $r-i\sim0.9$, and $i-z\sim0.45$, which represent also the areas of highest density in the colour-colour plots of Fig.~\ref{fig:paper2SEDfitcolcolbc03expbeta5}.\\

\begin{figure}
  \centering
  \includegraphics[width=0.9\columnwidth]{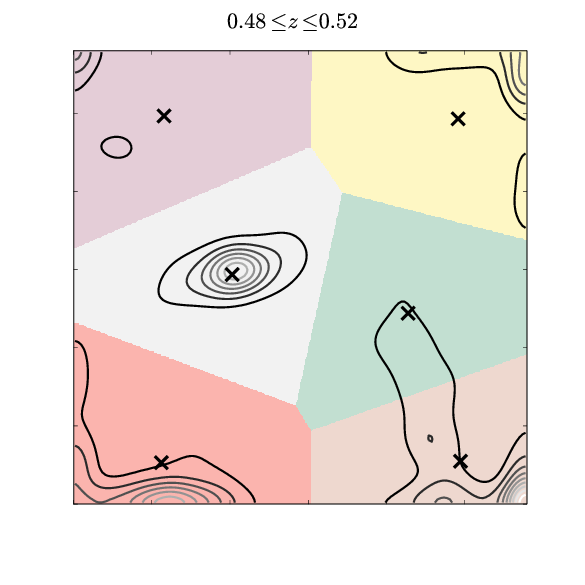}
  \caption{
    $k$-means clustering of the \gls{SOM} for $0.48\leq z\leq0.52$ from Fig.~\ref{fig:SOM5}.
    The Voronoi bins are highlighted by different colours, and cluster centres are marked by black crosses.
    Contours show the number frequency of the data (note that the contours are smoothed to increase the clarity of the figure) and are drawn at $10\,\%$ intervals between $10\,\%$ (black) and $90\,\%$ (white).
  }
  \label{fig:km5}
\end{figure}

As previously mentioned, we aim to select a sample of model \glspl{SED} from the best fitting models of the last section, where we want to take these density variations in the population  of the data into account.
Therefore, we want to identify clusters in the mapped data to select a model from each cluster that should represent the galaxies within the same cluster cell in terms of colour and absolute $r$ band magnitude.
To perform the cluster search we employ a $k$-means clustering algorithm \citep{zbMATH03129892} that partitions the two-dimensional maps into cluster cells.
Each cell is a Voronoi bin \citep{voronoi}, where two bins or clusters are separated from one another by a border orthogonal to the line connecting the cluster centres.
Every data point belongs to the cluster with the nearest centre.
Specifically, we use the python $k$-means clustering algorithm included in the \ttt{scikit-learn} package \citep{scikit-learn}.\\
Fig.~\ref{fig:km5} presents the bins of the $k$-means clustering algorithm with $k=6$ clusters applied to the \gls{SOM} of Fig.~\ref{fig:SOM5}, where the cluster centres are indicated by black crosses.
The density of the data in the \gls{SOM} is shown by contours in the plot.
The data exhibit a number of clusters in all redshift bins which does not exceed six, wherefore $k=6$ was chosen.\\
We want to select a set of \glspl{SED} from the best fitting model \glspl{SED} which we will use in the following as template set for \glspl{photoz}.
Therefore, for each cluster bin we take all models into account whose corresponding data points also lie in the same bin.
Then, for each of these models separately, we estimate photometric redshifts on all objects within the cell and determine resulting quality parameters, i.e., mean error, scatter and catastrophic failures of the \glspl{photoz}.
For each cluster we chose a model that yields the best \gls{photoz} results in terms of these parameters.
Thereby, we have for each of the four redshift bins six models, hence $24$ models in total.
The such selected \glspl{SED} will be used in the following to estimate \glspl{photoz}.


\section{Photometric Redshifts}
\label{sec:BOSSphotozresults}
In this section we analyse the photometric redshift results we get using our novel templates in combination with the \cphotoz\ code on the whole CMASS sample detailed in Sec.~\ref{sec:paper2BOSSCMASSsample}.
Afterwards, we compare our \glspl{photoz} with the photometric redshifts provided by the \gls{SDSS} database.


\subsection{Photometric Redshifts with the Novel Template SEDs}
\label{sec:paper2novelphotozs}
\begin{figure*}
  \centering
  \includegraphics[width=0.8\textwidth]{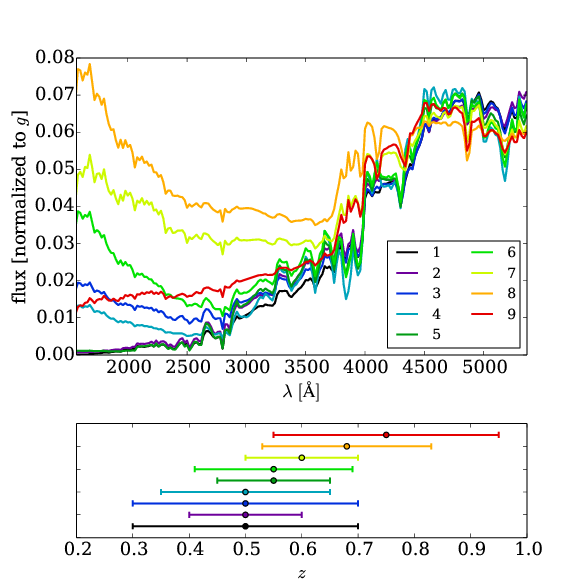}
  \caption{
    Upper panel: The nine surviving model \glspl{SED} within the range of the \gls{SDSS} $ugriz$ filter system at redshifts $0.45\leq z \leq0.9$ (the $z$ range of our galaxy sample).
    Lower panel: Corresponding $z$ prior parameters, $\mu_z$ and $\sigma_z$.
    The colour code is the same as in the upper panel.
  }
  \label{fig:paper2SEDs}
\end{figure*}

\begin{figure*}
  \centering
  \includegraphics[width=0.7\textwidth]{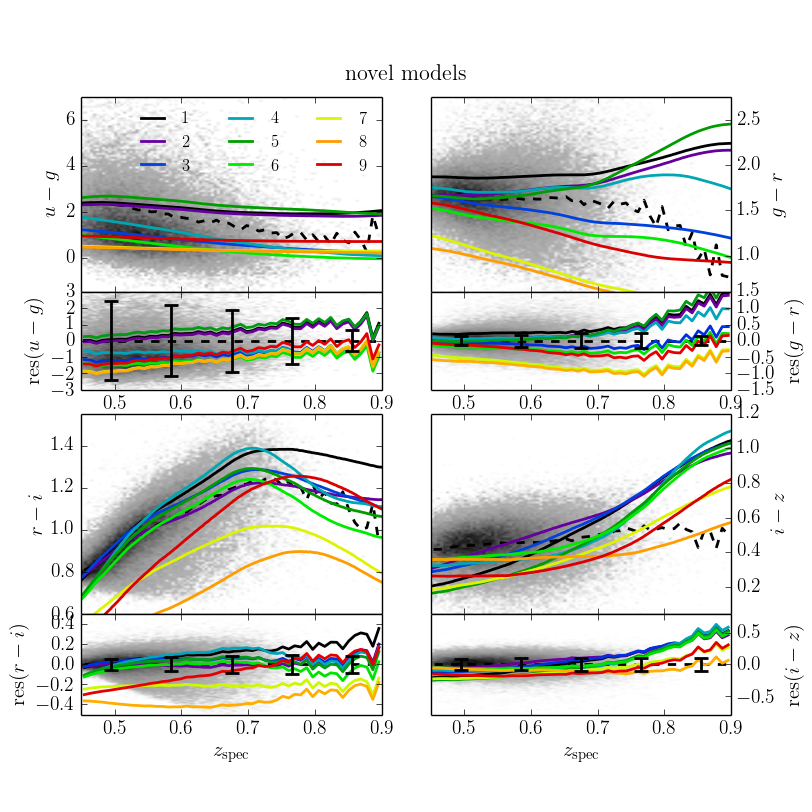}
  \caption{
    In the upper panels of the four major panels the CMASS colours are plotted as a function of redshift in grey, and the median is shown by a black dashed line.
    The lower panels are normalised to the median colour and show the residuals.
    Error bars present the median data errors in five redshift bins.
    On top of that the predicted colours of the nine selected models are drawn.
    The colour code matches that of Fig.~\ref{fig:paper2SEDs}.
  }
  \label{fig:paper2zvscolSOM}
\end{figure*}

To introduce the models created in the previous section into the \cphotoz\ code, we define the $z$ prior such that $\mu_z$ of a model \gls{SED} is the central value of the $z$ bin of the catalogue from which the model originates, wherefore $\mu_z=0.5,0.6,0.7,0.8$.
As $\sigma_z$, we set a default value of $0.2$ which leads to a prior $P_z(z|T)$ function wide enough to avoid focusing effects at certain redshift bin centres, while ensuring a smooth transition between them.
The resolution in redshift of a \gls{photoz} run is $0.002$ in the range of $z\in[0.0,3.0]$.
The allowed redshift range is much wider than that which is populated by the CMASS galaxies, such that we can analyse if the \gls{photoz} accuracy is diminished by $\mzp$ values that are highly over- or underestimated.
This is done because we would like to be able to run \gls{photoz} codes with the new models on catalogues with galaxies from larger redshift ranges and with more variations in \gls{SED} type.
We therefore want to make sure that small errors in $\mzp$ are not due to a restriction in the redshift range.\\
To improve the priors by adapting them iteratively, we analyse the outcome for subsamples of objects which are fitted best by a specific model.
Thereby we can adjust the redshift and luminosity priors for each model \gls{SED} in order to reduce outliers and bias.
Essentially, we decrease the value of $\sigma_z$ whenever a model yields lower $\mzp$ accuracies for redshifts further away from its bin centre.
Furthermore, we allow $\mu_z$ to vary if the \gls{photoz} performance of a specific model can be enhanced.
If we observe that a specific template provides very bad redshifts which cannot be resolved by adjusting the respective prior, we omit these models completely in following runs.
This is mostly the case for models created from the highest two redshift bins.
Since we chose the models only on account of their \gls{photoz} performance on a redshift bin, they might still yield a bad estimate in redshift ranges outside the bin.
The resulting model set then consists of nine \glspl{SED} with adapted redshift priors.
The luminosity priors where set initially to $\mu_M=20.0$, $\sigma_M=6.0$, and $p_M=6.0$, to allow for a wide range of higher luminosities.
The high exponent leads to a very flat functional behaviour within $\sigma_M$, and to steep decreases in $P_M(M)$ at $\mu_M\pm\sigma_M$.
Adjusting $P_M(M)$ in the iteration is not necessary, since we cannot detect outliers which could be avoided through a different luminosity prior.\\
The model \glspl{SED} are shown in Fig.~\ref{fig:paper2SEDs}, where we plot them in the wavelength range covered by \gls{SDSS} at the redshifts of our catalogue $0.45\leq z\leq 0.9$.
In the lower panel we present the redshift prior parameters $\mu_z$, and $\sigma_z$ which correspond to the model \glspl{SED}.
Fig.~\ref{fig:paper2zvscolSOM} shows the colours of the nine models as a function of redshift.
With the nine models we can account for the large spread of the data in most cases.
Furthermore, we cover also the bluer parts in \gmr, \rmi, and \imz\ (cf. Fig.~\ref{fig:BOSSzcolM09}).
For lower redshifts, our models produce colours that cover only the bluer \rmi\ and \imz\ ranges of the data.
This is not because the \bcexpb\ do not fit the colours of the data (cf. Figs.~\ref{fig:paper2SEDfitcolcolbc03expbeta5} to \ref{fig:paper2SEDfitcolcolbc03expbeta8}), but accidental, since the models were selected (from within their cluster bins) on account of their \gls{photoz} performance in Sec.~\ref{sec:BOSSselection}.
The model SEDs with designations $7$ to $9$ cover only peculiar blue colours of the data.
These models were created on the basis of the higher $z$ samples, which is why they match the data better at higher redshifts (especially the \imz\ colour).
The bluer colours of the models can only be observed from a small number of galaxies.
Therefore, when all models are fitted to the spectroscopic redshift (but also in the \gls{photoz} estimation below), the models $7$-$9$ are best fitting only for $\sim1$ to $5\,\%$ of galaxies.\\
The physical parameters and the $\lambda_i$ and $\beta$ values of the nine model \glspl{SED} are summarised in Tab.~\ref{tab:paper2physparam}.
They explain the behaviour of the \glspl{SED}.
From Fig.~\ref{fig:paper2SEDs} we see that the model \glspl{SED} roughly follow a trend and become bluer with increasing redshift (except for the red coloured \gls{SED} with $\mu_z=0.8$), which is mirrored in the values of Tab.~\ref{tab:paper2physparam}.
The $e$-folding time scale $\tau$ roughly increases, as well as the burst fraction, making the resulting \glspl{SED} bluer on average.
The red highlighted \gls{SED} has lower fluxes than the orange and the yellow-green \glspl{SED} in the UV part of the spectrum, not continuing the sequence.
It is redder because of the high extinction values and because of its (relatively) high metallicity, $Z=\mZsun$.\\
In summary, although we can see a qualitative trend in the \glspl{SED} as a function of redshift prior (which originates from the spectroscopic redshifts of the underlying $z$ bin with small adaptions), the trends in the physical parameters are not that evident.
This is because they are degenerate and changes in one parameter can yield similar results in the SEDs as a variation in another parameter (e.g., the well-known age-metallicity degeneracy).

\begin{table*}
  \centering
  \caption{
      Physical parameters of the nine surviving templates.
      The first column gives the numbering of the models and the plot colour according to Figs.~\ref{fig:paper2SEDs} and \ref{fig:paper2zvscolSOM}.
      The column ''burst`` is the mass fraction of the burst, whereas $A_{V,\,\mathit{burst}}$ is the burst extinction.
  }
  \begin{tabular}{l|ccccccccc}
    \textbf{model \#}  & $\mathbf{Z}$   & $\boldsymbol{\tau\,[}\mathbf{Gyr}\boldsymbol{]}$ & \textbf{age}\,$\boldsymbol{[}\mathbf{Gyr}\boldsymbol{]}$ & $\mathbf{A_V}$ & \textbf{burst} & $\mathbf{A_{V,\,\mathbf{burst}}}$ & $\boldsymbol{\lambda_\mathbf{i}}\,\boldsymbol{[}\mathbf{\AA{}}\boldsymbol{]}$ & $\boldsymbol{\beta}$\\
    \hline
    1 (black)          & 0.008 & SSP            & 2.0       & 1.2   & \textendash & \textendash & 5000 & 2.0 \\
    2 (violet)         & 0.004 & SSP            & 6.0       & 0.6   & 1.0\,\%     & 2.0         & 5500 & 1.5 \\
    3 (blue)           & 0.008 & 1.0            & 6.0       & 1.0   & 1.0\,\%     & 0.0         & 5000 & 1.5 \\
    4 (turquoise)      & 0.05  & SSP            & 8.0       & 0.7   & 1.0\,\%     & 0.0         & 3500 & 2.0 \\
    5 (dark green)     & 0.02  & SSP            & 4.0       & 0.0   & 1.0\,\%     & 1.0         & 5000 & 1.5 \\
    6 (green)          & 0.05  & SSP            & 3.0       & 1.3   & 2.0\,\%     & 0.0         & 3500 & 2.0 \\
    7 (yellow-green)   & 0.05  & 3.0            & 5.0       & 1.4   & 1.0\,\%     & 0.0         & 5000 & 2.0 \\
    8 (orange)         & 0.004 & 50.0           & 4.0       & 2.7   & 1.0\,\%     & 0.0         & 3000 & 1.5 \\
    9 (red)            & 0.02  & 3.0            & 4.0       & 2.2   & 2.0\,\%     & 2.0         & 4500 & 2.0
  \end{tabular}
  \label{tab:paper2physparam}
\end{table*}

Before analysing the \gls{photoz} performance of the new models, we want to introduce several metrics which we use to assess the photometric redshift quality.
We decide to provide a large number of metrics to enable the reader to compare with other publications.
The photometric redshift error is $\dz=\mzp-\mzs$, and $\dzrf=\Delta z/(1+\mzs)$ in the rest frame.
Catastrophic outliers are defined such that $|\dzrf|\geq0.15$ \citep[cf.][]{2006A&A...457..841I}.
The mean errors are characterized by the bias $\langle\dz\rangle$ and the mean absolute error $\langle|\dz|\rangle$, as well as by their corresponding values in the rest frame, $\langle\dzrf\rangle$ and $\langle|\dzrf|\rangle$.
The root of the sample variance is denoted $\sdzrf$, and $\ssixtyeightrf$ is the half of the width of the $\dzrf$ distribution where $68.27\,\%$ of the sample is located, corresponding to a $1\sigma$ confidence interval.
Finally, in terms of scatter we also calculate the normalised median absolute deviation $\sigma_{\mathrm{NMAD}}\equiv1.48\cdot\mathrm{median}(|\dzrf|)$ \citep{2006A&A...457..841I} which is calculated for non-outliers only, and gives a clue about the width of the $\dzrf$ distribution without regarding the flanks.
We calculate the fractions $\eta_{2\sigma}$, $\eta_{3\sigma}$ of galaxies within $\langle\dzrf\rangle\pm i\cdot\sdzrf$, where $i=2,3$ \citep[cf.][]{2014MNRAS.438.3409C}.
If $\dzrf$ behaved as a perfect Gaussian, then $2\cdot\sdzrf$ and $3\cdot\sdzrf$ would be the $2\sigma$ and $3\sigma$ confidence intervals and therefore hold $95.45\,\%$ and $99.73\,\%$ of the objects.
But the distribution of $\dzrf$ can be non-Gaussian and still yield reasonable values for $\eta_{2\sigma}$ and $\eta_{3\sigma}$ (compare Figs.~\ref{fig:noveldz} and \ref{fig:BOSSRFdz} later on).\\
To evaluate the precision of the \gls{photoz} errors $\delta z$ estimated by the code (Eq.~\ref{eqn:paper2PhotoZerrors}) from the \gls{PDF} (cf. Sec.~\ref{sec:paper2PhotoZ}), we introduce $\Delta z^\prime\equiv\dz/\delta z$.
If the errors are estimated correctly, the distribution of $\dz^\prime$ should resemble a standard normal distribution \citep[cf.][]{2014MNRAS.445.1482S}.
We will therefore analyse the values of the mean $\langle\dz^\prime\rangle$ and the standard deviation $\sigma_{\dz^\prime}$ of the distribution of $\dz^\prime$.\\
Lastly, we perform Kolmogorov-Smirnov (KS) tests on the distributions $N(\mzp)$ and $P(\mzp)$ \citep[cf.][]{2014MNRAS.438.3409C}, where the latter is the $\mzp$ distribution derived by stacking the \glspl{PDF} for all objects.
The KS test value is the maximum absolute difference between the cumulative distribution functions $\int_{-\infty}^{z}P(\hat{z})d\hat{z}$ of the probability density of $N(\mzp)$, or the stacked $P(\mzp)$, to the cumulative distribution of $N(\mzs)$.\\

\begin{figure*}
  \centering
  \includegraphics[width=0.7\textwidth]{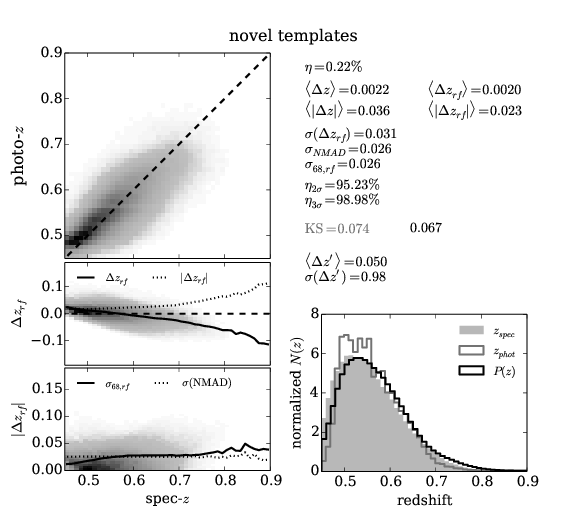}
  \caption{
    Photo-$z$ results with novel templates, based on the BC03 $\lambda^{-\beta}$ models and selected through a \gls{SOM} and a $k$-means clustering algorithm.
    \emph{Left Panels:} The upper panel shows the spectroscopic versus the photometric redshift, where the dashed line is at $\mzp=\mzs$.
    The middle panel is the distribution of the rest frame \gls{photoz} error as a function of the spectroscopic redshift, where we plot the bias (solid line) as well as the mean absolute error (dashed line). 
    Finally, the lower panel shows $\mzs$ versus the absolute rest frame error, where we plot $\sigma_{68}$ (solid) and $\snmad$ (dashed) on top.
    \emph{Right Panel:} The relative frequency distributions of $\mzs$ (grey filled histogram), $\mzp$ (dark grey line), and the stacked \gls{PDF} of all objects (black line).
    Additionally, the \gls{photoz} quality metrics are printed in the figure.
    We provide two values for the $\mathit{KS}$ test, for the predictions of single $\mzp$ values (from the mode of the \gls{PDF}, grey), and for the \gls{PDF} (black).}
  \label{fig:novel}
\end{figure*}
After this introduction of the \gls{photoz} quality metrics we present the photometric redshift results of the CMASS sample with the novel template set and priors of Sec.~\ref{sec:BOSSselection} and Fig.~\ref{fig:paper2SEDs} in Fig.~\ref{fig:novel}.
The upper left panel shows $\mzs$ versus $\mzp$, the middle left panel shows the photometric redshift rest frame error as a function of \sz, and the lower left panel presents $\mzs$ versus $|\dzrf|$.
We indicate the median $\dzrf$ and $|\dzrf|$ values in the middle left panel by solid and dashed black lines, whereas in the lower panel we highlight the $\ssixtyeightrf$ and $\snmad$ values as a function of the spectroscopic redshift also by solid and dashed black lines.
Finally, the right panel presents the normalised redshift distributions for $\mzs$ (grey filled histogram), the distribution derived from the single-value photometric redshifts $z_\mathrm{mode}$ (grey line) derived by the mode of $P(z)$, and the stacked \glspl{PDF} (black line).
The photometric redshift quality metrics discussed above are printed in the plot as well.\\
The bias has a small positive value for the lowest considered redshifts, then is close to zero, and is decreasing to negative values for higher redshifts $z\sim0.7$.
The overall mean value is still positive due to the small sample size at higher redshifts, visible by the number density in Fig.~\ref{fig:paper2zmabszspec} or the right panel of Fig.~\ref{fig:novel}.
The scatter (lower left panel in Fig.~\ref{fig:novel}) increases slightly with increasing \sz, while the value of $\snmad$ stays more or less the same.
This means, that the outliers (which are not considered in the calculation of $\snmad$) are predominantly responsible for an increase in $\sdzrf$.
The fraction of catastrophic outliers however is very small $\eta=0.22\,\%$.
In the right panel of Fig.~\ref{fig:novel} we observe that the \gls{photoz} predictions from the mode of the \gls{PDF} yield deviations from $N(\mzs)$ larger than for the case where the whole \glspl{PDF} are considered.
The excesses observed in $N(\mzp)$ are due to the overestimation mentioned previously.
They are not anymore visible when we use the posterior $P(z)$ distributions in the $N(z)$ reconstruction, which means that the \gls{PDF} should be favoured for science analyses \citep[cf., e.g.,][]{2008MNRAS.386..781M,2012MNRAS.421.2355H}.\\
As for the error estimation of the \cphotoz\ code, we see from the values of $\langle\dz^\prime\rangle$ and $\sigma_{\dz^\prime}$, that $\dz^\prime$ behaves very close to a standard normal distribution, which is the goal in the photometric redshift error predictions.
This means that not only the approach for the calculation of $\delta z$ is legitimate, but also that the models and priors create a reasonable $P(z)$ (from which $\delta z$ is extracted).
\\
Tab.~\ref{tab:metrics} in Sec.~\ref{sec:paper2SDSSresults} holds a summary of the derived quality metrics, together with results of public \glspl{photoz} from the \gls{SDSS}, which we will analyse in Sec.~\ref{sec:paper2SDSSresults}.\\
To investigate if the estimated errors are reliable in identifying outliers, we assume the null hypothesis that an object is a \gls{photoz} outlier.
The probability of an outlier being falsely classified as a non-outlier on account of $\delta z$ is then the type I error $\alpha$.
Additionally, the type II error $\beta$ gives the probability of a non-outlier being misclassified as an outlier by the estimated error.
For the \glspl{photoz} of this work we get $\alpha=0.93$ and $\beta=3.5\cdot10^{-4}$.
This means that an outlier is falsely classified as a non-outlier with probability $\alpha$, but also that a non-outlier is almost never classified as an outlier ($\beta$).
Although $\sigma(\Delta z^\prime)$ is close to, but slightly smaller than, $1$, the deviation is probably caused by to shallow peaks around the mode of $P(z)$ for outliers, such that their errors are underestimated.
The $\alpha$ and $\beta$ value are summarised together with the results of the \gls{SDSS} (Sec.~\ref{sec:paper2SDSSresults}) in Tab.~\ref{tab:paper2significance}.

\begin{figure} 
  \centering
  \includegraphics[width=\columnwidth]{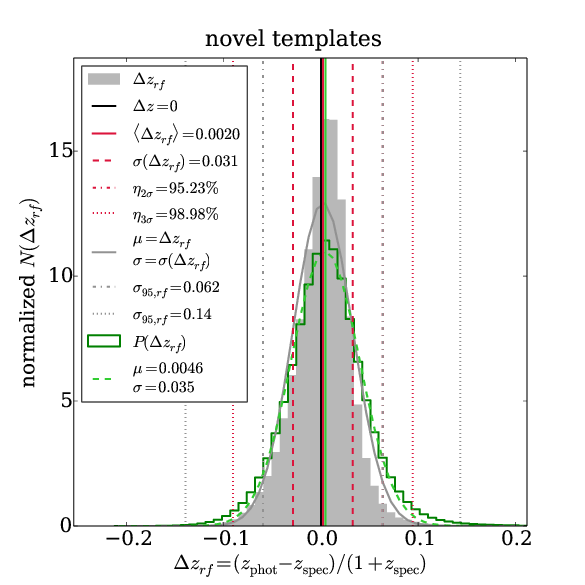}
  \caption{
    Photometric redshift rest frame error distribution (grey filled histogram) estimated with the novel templates and priors.
    $\dz=0$ is highlighted by a solid black line, whereas the bias is shown by a solid red line.
    The steps at $\dzrf\pm\sdzrf$, $2\sdzrf$, and $3\sdzrf$ are represented by dashed, dash dotted, and dotted red lines, and the corresponding number fractions are given in the legend.
    We also plot a Gaussian with $\mu=\langle\dzrf\rangle$, and $\sigma=\sdzrf$ in grey.
    Furthermore, dash dotted and dotted grey vertical lines show the real $\stwosrf$ and $\sthreesrf$ intervals.
    Since $\stwosrf\approx2\sdzrf$ we cannot discern the red and the grey dash dotted lines.
    The $\dzrf$ derived from the stacked \glspl{PDF} is shown by a green solid histogram, to which we fit a Gaussian highlighted by a dashed green curve.
  }
  \label{fig:noveldz}
\end{figure}
The distribution of the photometric redshift errors in the rest frame is shown in Fig.~\ref{fig:noveldz} by a grey filled histogram.
In red we indicate the bias and the ranges of $\langle\dzrf\rangle\pm i\cdot\sdzrf$, where $i=1,2,3$.
Furthermore, we calculate the real interval widths where $95.45\,\%$, and $99.73\,\%$ of the galaxies are located and introduce them in the plot (black dash dotted and dotted vertical lines).
Finally, we calculate the $\dzrf$ distribution from the stacked \glspl{PDF} and fit a Gaussian to the histogram (green lines).\\
The peak of the $\dzrf$ distribution is slightly shifted to the right, which is again due to the overestimated \glspl{photoz} at the lowest redshifts.
Concerning the resemblance to a Gaussian, we observe from the values given in the plot that $\stwosrf\approx2\sdzrf$ (which is why we cannot discern the lines in Fig.~\ref{fig:noveldz}) and $\sthreesrf>3\sdzrf$.
The latter is due to the outliers which can be scattered far from the \sz\ value due to the allowed fitting range.
The fractions $\eta_{2\sigma}$ and $\eta_{3\sigma}$ of objects within $2\sdzrf$ and $3\sdzrf$ have reasonable values.
While $\eta_{2\sigma}=95.23\,\%$ is very close to the desired value of $95.45\,\%$, $\eta_{3\sigma}=98.98\,\%$ is slightly smaller by only ${\sim}\,0.8\,\%$.
The $\dzrf$ distribution derived from the \glspl{PDF} is slightly broader than the distribution of the individual \gls{photoz} results.
This is a consequence of the asymmetry of the \glspl{PDF} that often have higher probabilities for redshifts greater than the most probable $z$.\\
\begin{figure} 
  \centering
  \includegraphics[width=\columnwidth]{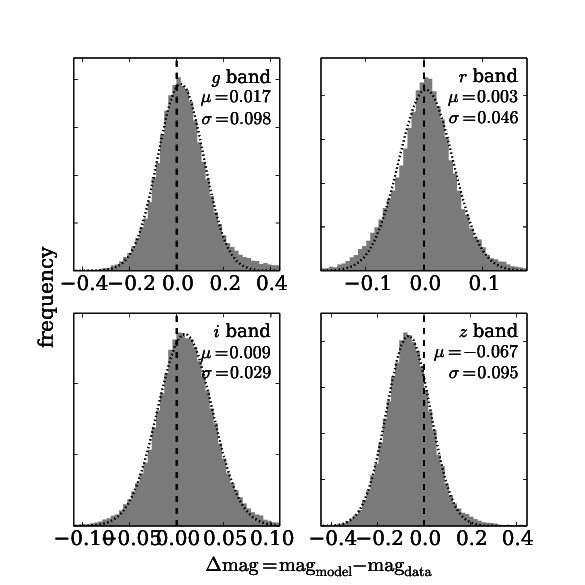}
  \caption{
    Deviations in magnitudes of the predicted magnitudes of the novel model \glspl{SED} to the data.
  }
  \label{fig:novelmag}
\end{figure}
Fig.~\ref{fig:novelmag} presents the deviations of magnitudes, predicted by the models in a \gls{photoz} run, and the data.
A Gaussian is fitted to the histograms whose best fitting parameters are printed in the panels.
For the $r$ and $i$ band magnitudes the $\Delta\mathrm{mag}$ distributions are very narrow, while the expectation values of the Gaussians are near zero.
This is thanks to the depth of the photometry in these filters, which is greatest in $i$ and $r$, and the photometry has the smallest measurement uncertainties.
For the $g$ and the $z$ band magnitudes however, the broadness of the $\Delta\mathrm{mag}$ distributions increases, which is mostly an effect of the more shallow photometry in these bands, but also of the respective magnitude ranges of the galaxies.
Furthermore, the mean magnitude deviations in these two bands are higher than in $r$ and $i$.
While $\mu$ of the best fitting Gaussian is still small in comparison to $\sigma$ for the $g$ band, it is relatively high in the $z$ band.
We selected the models for the \gls{photoz} estimation only on account of their \gls{photoz} performance, and not on how well they match the data in terms of magnitudes.
However, this does not necessarily imply that the \bcexpb\ models are not able to fit the data in terms of magnitudes.
Indeed, calculating the average magnitude offsets of the \gls{SED} fitting of Sec.~\ref{sec:paper2SEDfittingresults}, yield maximum values of $[0.6,0.9,0.3,1.1]\cdot10^{-3}$ in $g$, $r$, $i$, and $z$ for the four redshift bins.\\
In the next section we want to compare the photometric redshift results of this section with the \glspl{photoz} provided by \gls{SDSS}.


\subsection{Comparison to SDSS Photometric Redshifts}
\label{sec:paper2SDSSresults}
\begin{figure*}
  \centering
  \includegraphics[width=0.7\textwidth]{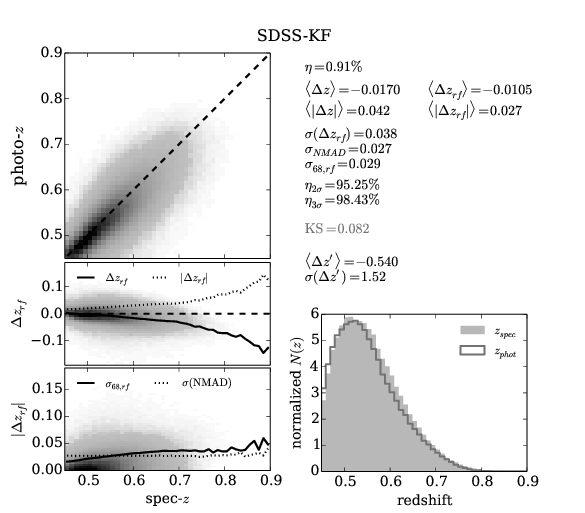}
  \caption{
     Photo-$z$ results from the KF approach published by \gls{SDSS}.
     See Fig.~\ref{fig:novel} for a detailed description.
  }
  \label{fig:BOSSNNzszp}
\end{figure*}
\begin{figure} 
  \centering
  \includegraphics[width=\columnwidth]{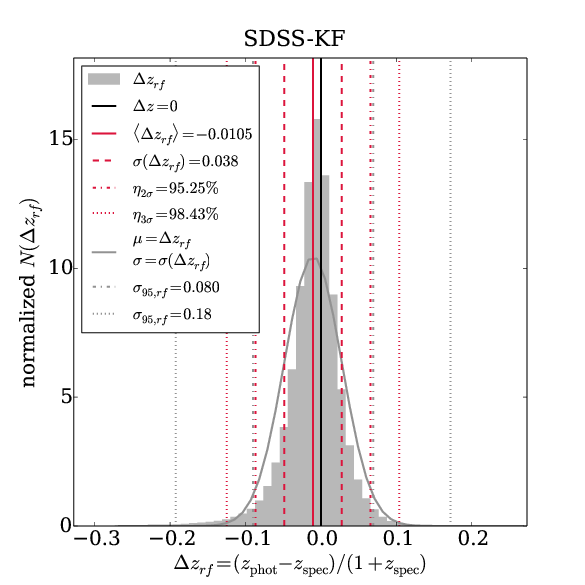}
  \caption{
    Photometric redshift rest frame error distribution (grey filled histogram) estimated by the SDSS-KF method.
    A detailed explanation is given in Fig.~\ref{fig:noveldz}.
  }
  \label{fig:BOSSNNdz}
\end{figure}
We present the photometric redshifts from the \gls{SDSS}-III database in comparison to the results of our code-template combination.
The \gls{SDSS} database provides \glspl{photoz} from two different empirical methods.
One code uses a $k$-d tree nearest neighbour fit to derive the redshifts \citep[hereafter KF after][]{2007AN....328..852C}.
The SDSS-KF results for our catalogue are shown in Fig.~\ref{fig:BOSSNNzszp}, which is equivalent to Fig.~\ref{fig:novel} and also shows the quality metrics.
From the middle panel of Fig.~\ref{fig:BOSSNNzszp} we see that the location of the densest part of the population lies close to the $\dz=0$ line for low redshifts $z\lesssim0.6$.
In all higher $\mzs$ regions the photometric redshift is systematically underestimated which leads to the low negative value in the bias.
An underestimation is also present in the results from Sec.~\ref{sec:paper2novelphotozs}, but it is slightly higher in the SDSS-KF case.
From the lower panel of Fig.~\ref{fig:BOSSNNzszp} we see that also here the rise in $\ssixtyeightrf$ is mostly caused by outliers, which are not considered in the calculation of $\snmad$ (which stays more or less constant).
Taking the estimated photometric redshift errors into account, the relatively high bias does not decrease, which is mirrored in the higher absolute value of $\langle\dz^\prime\rangle$, but even still increases.
Hence, the bad estimates of the \glspl{photoz} are not recognized by the error estimation.
This can also be seen in the value of $\sigma_{\dz^\prime}$ which is well beyond $1$ and means that the errors are on average underestimated.
This also affects the results of a significance test on outlier classification (cf. Sec.~\ref{sec:paper2novelphotozs}).
The errors type I and II read (at two significant figures) $\alpha=1.0$ and $\beta=7.9\cdot10^{-4}$ (see also Tab.~\ref{tab:paper2significance} for comparison with the other results), confirming that the errors are underestimated.\\
In terms of $N(\mzp)$ the low \gls{photoz} values at higher $\mzs$ do not change the shape significantly, which is due to the low sample size in these regions, and results in the low value of \emph{KS}.
We cannot compare to the $N(z)$ results for \glspl{PDF}, since they are not provided.\\
As before, we show in Fig.~\ref{fig:BOSSNNdz} the $\dzrf$ distribution.
Comparing with our case from Fig.~\ref{fig:noveldz}, we see that here the results are slightly shifted to the left, an effect due to the underestimated \glspl{photoz} at higher redshifts.
Also in the case of the SDSS-KF redshifts, the $\stwosrf$ and $\sthreesrf$ values do not perfectly coincide with $2\sdzrf$ and $3\sdzrf$, but are elevated.
$\eta_{2\sigma}=95.25\,\%$ is very close to the desired value, while $\eta_{3\sigma}=98.43\,\%$ deviates by ${\sim}\,1.3\,\%$.\\
In comparison to the redshifts we get with our novel templates and code, the quality metrics for bias, error, and scatter are all higher in the case of the SDSS-KF \glspl{photoz}.
Furthermore, the estimated redshift errors are more reliable in our code-template combination.
The KS test yields better results in the case of this work, since the SDSS-KF \glspl{photoz} are systematically underestimated.\\
\begin{figure*}
  \centering
  \includegraphics[width=0.7\textwidth]{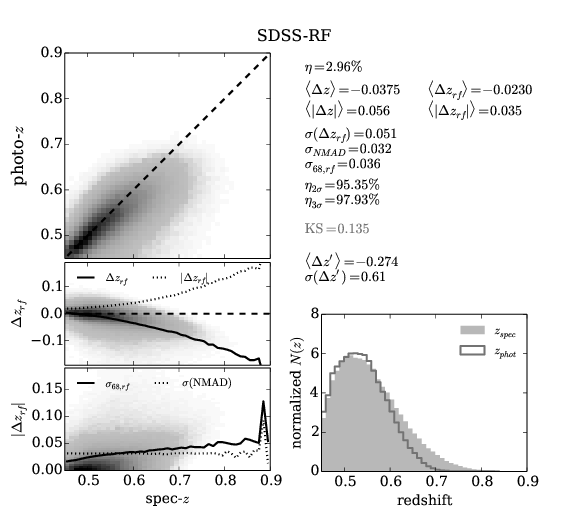}
  \caption{
    Photo-$z$ results from the RF approach published by the \gls{SDSS}.
    See Fig.~\ref{fig:novel} for a detailed description.
  }
  \label{fig:BOSSRFzszp}
\end{figure*}

\begin{figure} 
  \centering
  \includegraphics[width=\columnwidth]{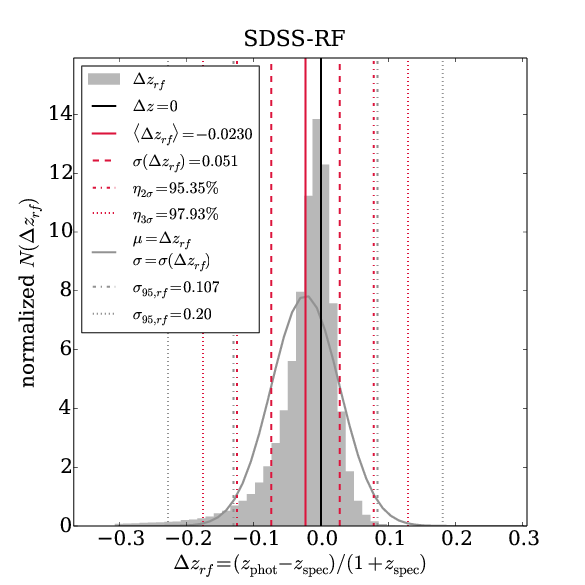}
  \caption{
    Photometric redshift rest frame error distribution of the SDSS-RF published by the \gls{SDSS}.
    A detailed explanation is given in Fig.~\ref{fig:noveldz}.
  }
  \label{fig:BOSSRFdz}
\end{figure}

The second photometric redshift results published by the \gls{SDSS} are from another empirical code, which uses random forests to predict \glspl{photoz} \citep[][hereafter SDSS-RF]{2010ApJ...712..511C}.
Fig.~\ref{fig:BOSSRFzszp} presents the results of the SDSS-RF code.
In this case, the outlier rate is significantly higher than in both previous cases, as well as the mean and the absolute rest frame errors.
Above $z\sim0.55$ the \glspl{photoz} are systematically underestimated to a greater extent than in the former two cases, yielding the low bias.
Furthermore, the scatter values $\sdzrf$, $\ssixtyeightrf$, and $\snmad$ have also increased in comparison to the results of this work, as well as compared to the SDSS-KF redshifts.\\
In the distribution of $N(\mzp)$ we can observe the lack of higher \gls{photoz} values which accumulate around $z\sim0.55$.
Because of that, the KS test value is higher.
In terms of their estimated photometric redshift errors, the SDSS-RF method outperforms the SDSS-KF approach, but still produces a larger bias in $\langle\dz^\prime\rangle$.
The errors are furthermore overestimated, visible in the $\sigma_{\dz^\prime}$ value which is lower than the desired value of $1$.
Again, this affects also the outlier classification derived from $\delta z$.
In the case of SDSS-RF \glspl{photoz}, the null hypothesis of an object being an outlier yields $\alpha=0.86$ and $\beta=0.015$ (shown together with the previous results in Tab.~\ref{tab:paper2significance}).
The higher values of $\delta z$ yield a smaller probability of an object being misclassified as a non-outlier, but also raises $\beta$ (although only to a small value).
So although the errors are overestimated on average, outliers are still misclassified at a significance level of $0.86$.\\
In Fig.~\ref{fig:BOSSRFdz} we plot the $\dzrf$ distribution.
The shift to the left is present to an even greater extent than in the previous case, and strongly deviating from a Gaussian with $\mu=\langle\dzrf\rangle$ and $\sigma=\sdzrf$ also plotted in Fig.~\ref{fig:BOSSRFdz}.
This is a result of the underestimation of the \glspl{photoz}.
Furthermore, as in the previous cases, the value of $\stwosrf$ is marginally higher than $2\sdzrf$, while the deviation is greater in $\sthreesrf$.
On the other hand $\eta_{2\sigma}=95.35\,\%$ is even closer to $95.45\,\%$ than in both previous cases, while $\eta_{3\sigma}=97.93\,\%$ deviates more strongly, ${\sim}\,1.8\,\%$.\\

\begin{table*}
  \centering
  \caption{
    Summary of photometric redshift quality metrics of the template fitting results with the novel templates and priors used with the \cphotoz\ code, the \gls{SDSS} \glspl{photoz} of the KF code, and the random forest code (SDSS-RF).
    Values for \emph{KS} in the case of this work are derived from $P(z)$, whereas the values in brackets are calculated using only the $z_\mathrm{mode}$ predictions.
  }
  \begin{tabular}{l|cccccccccc}
    \textbf{setup}    & $\boldsymbol{\eta\,[\%]}$
    & $\mathbf{\boldsymbol{\langle\Delta z_{rf}\rangle}}$
    & $\mathbf{\boldsymbol{\langle|\Delta z_{rf}|\rangle}}$
    & $\boldsymbol{\mathbf{\sigma (\Delta z_{rf})}}$ & $\boldsymbol{\sigma}_{\mathrm{\mathbf{NMAD}}}$
    & $\boldsymbol{\mathbf{\sigma}_{68,\mathbf{rf}}}$
    & $\boldsymbol{\mathbf{\langle\Delta z^\prime\rangle}}$
    & $\mathbf{\boldsymbol{\sigma_{\dz^\prime}}}$ & \textbf{KS}\\ 
    \hline
    this work           & $0.22$ & $0.0020$  & $0.023$    & $0.031$  & $0.026$    & $0.026$   & $0.050$  & $0.98$    & $0.067\,(0.074)$\\
    SDSS-KF  & $0.91$ & $-0.011$  & $0.027$    & $0.038$  & $0.027$    & $0.029$   & $-0.54$  & $1.5$     & $0.082$\\
    SDSS-RF      & $3.0$  & $-0.023$  & $0.035$    & $0.051$  & $0.032$    & $0.036$   & $-0.27$  & $0.61$    & $0.14$
  \end{tabular}
  \label{tab:metrics}
\end{table*}
\begin{table}
  \centering
  \caption{
    Significance test for outlier classification.
    The null hypothesis is that an object is an outlier, and $\alpha$ and $\beta$ are the type I and II errors respectively (rounded to two significance figures).
  }
  \begin{tabular}{l|ccc}
                          & \textbf{this work}  & \textbf{SDSS-KF}   & \textbf{SDSS-RF}\\
    \hline
    $\boldsymbol{\alpha}$ & $0.93$              & $1.0$                         & $0.86$ \\
    $\boldsymbol{\beta}$  & $3.5\cdot10^{-4}$   & $7.9\cdot10^{-4}$             & $1.4\cdot10^{-2}$
  \end{tabular}
  \label{tab:paper2significance}
\end{table}
We present a summary of the photometric redshift quality metrics of the results from the template fitting of this work and the two \gls{SDSS} \glspl{photoz} in Tab.~\ref{tab:metrics}.
The model-prior combination of this work produces the lowest outlier fraction, bias, and mean absolute error compared to the \gls{SDSS} \gls{photoz} results.
When considered as a function of \sz\ the photometric redshifts derived in this work are slightly biased to lower $\mzp$ values for higher $\mzs$, but not to the same extent of the \gls{SDSS} codes.
Furthermore, the different scatter values calculated (i.e., $\sdzrf$, $\ssixtyeightrf$, and $\snmad$) are also lower for \glspl{photoz} of this work.
The distribution of redshift errors are more similar to a Gaussian for the models, priors, and code of this work, which can be observed in Figs.~\ref{fig:noveldz}, \ref{fig:BOSSNNdz}, and \ref{fig:BOSSRFdz}.
Considering the number fractions within $2\sdzrf$ and $3\sdzrf$ all three codes perform similarly well.
However, this does not mean that the $\dzrf$ distributions are necessarily good approximations of Gaussian distributions with $\mu=\dzrf$ and $\sigma=\sdzrf$ which can be observed in Figs.~\ref{fig:noveldz}, \ref{fig:BOSSNNdz} and \ref{fig:BOSSRFdz}.\\
When we evaluate the similarity between the redshift distributions, from which spectroscopic and photometric redshifts are sampled, through a KS test, the \glspl{photoz} of this work yield the best results for predictions of the mode which are improved when considering the whole $P(z)$  distribution.\\
The results of a significance test with the null hypothesis being that an object is a photometric redshift outlier is shown in Tab.~\ref{tab:paper2significance}.
Not one of the codes produces errors that can reliably predict outliers.
Albeit the SDSS-RF produces the best $\alpha$, this is mostly due to the overestimated errors (cf. Tab.~\ref{tab:metrics}).

\subsection{Deviations in Colour Predictions}
\label{sec:paper2colnovelNN}
In the last section we analysed the photometric redshift results with our code in comparison to the \gls{SDSS} \glspl{photoz}.
The redshifts from the SDSS-RF method were outperformed by those derived with the SDSS-KF code.
We want to analyse if the \glspl{photoz} of this work can produce the CMASS galaxy colours to a better extent than if we fit the same models to the data at the photometric redshifts of the SDSS-KF method.\\
\begin{figure*} 
  \centering
  \includegraphics[width=0.7\textwidth]{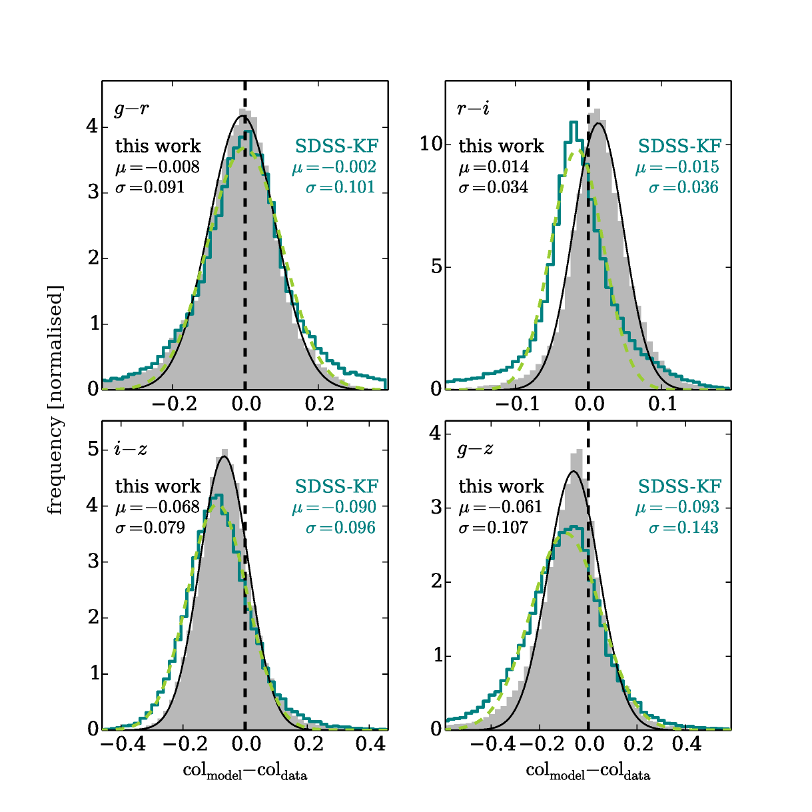}
  \caption{
    Deviation in \gmr, \rmi, \imz, and $g-z$ colours for photometric redshift results with the novel templates (grey filled histogram) and model predictions when fitting the same templates to the SDSS-KF \glspl{photoz} (blue histogram).
    Gaussians are fitted to both distributions, and are shown by a solid black line and a dashed green line respectively.
    The best fitting parameters of the Gaussians are printed in the panels.
  }
  \label{fig:dcolwethem}
\end{figure*}
Fig.~\ref{fig:dcolwethem} presents the differences between the model predictions and the data in the \gmr, \rmi, \imz, and $g-z$ colours.
The prediction of the \glspl{photoz} of this work are plotted in grey, the best fitting Gaussians are presented by black curves, and their parameters are given in the plot.
When fitted to the photometric redshifts of the SDSS-KF approach, the novel \bcexpb\ models from Sec.~\ref{sec:BOSSselection} predict colours which are represented by blue histograms in Fig.~\ref{fig:dcolwethem}, to which we also fit Gaussian curves.
In all four colours, the standard deviation of the $\Delta\mathrm{col}$ distribution is larger for SDSS-KF.
Furthermore, the mean deviation from the data colours is greater for SDSS-KF results in the \imz\ and $g-z$ colours.
$\mu$ is of the same order in \rmi\ for both cases.
It is smaller for SDSS-KF \glspl{photoz} in \gmr, but there both values are very low.
The fact that the SDSS-KF \rmi, \imz, and $g-z$ colours are shifted bluewards from the distribution in colours of this work is a result of the SDSS-KF \glspl{photoz} being more heavily underestimated.
All considered $\Delta\mathrm{col}$ distributions resemble their best fitting Gaussians well in the inner parts, with increased flanks.
However, this increase is more significant in for the SDSS-KF \glspl{photoz} in \rmi.\\
Therefore, the photometric redshifts derived in this paper not only generally have better quality metrics than the \gls{SDSS} \glspl{photoz}, but can furthermore recreate the colours of the CMASS galaxies to a better extent.


\section{Summary and Conclusions}
\label{sec:paper2summary}
In this work we created a set of model \glspl{SED} that are designed to match the colours of the \gls{BOSS} CMASS sample and provide accurate photometric redshifts.
We first analysed the colours of the \gls{LRG} model of \citet{2009MNRAS.394L.107M} which was created to match the \gmr\ and \rmi\ colours of the \gls{LRG} sample of \gls{SDSS}-II \citep{2001AJ....122.2267E}, and found that we cannot use a singular age or evolution configuration that matches the data in all colours.
Therefore, we created models for four redshift bins of widths $\Delta z=0.04$ centred on $\mzs=0.5,0.6,0.7,0.8$ with the stellar population synthesis code of \citet[BC03,][]{2003MNRAS.344.1000B}.
We generated \glspl{SSP} and \glspl{CSP} with decreasing \glspl{SFH} at various metallicities, and sampled the models from a wide age range.
These models were then fitted in superposition to a burst component and dust extinction to the data at known spectroscopic redshifts.\\
We observe a mismatch in the colours of the data and the models which is due to a too shallow red continuum slope in the model \glspl{SED}.
Large variations in the model parameters (i.e., metallicity, \gls{SFH}, age, burst, dust extinction, \gls{IMF}) do not produce models that better fit the data.
We speculate that inaccuracies in the modelling of the \gls{AGB} phase could be the reason for the observed colour mismatch.
In order to better recreate the colours of the CMASS galaxies we introduce additional degrees of freedom by modifying the model \glspl{SED} by multiplication with $\lambda^{-\beta}$ for $\lambda>\lambda_i$ with several values for $\lambda_i$ and $\beta$.
We showed that the BC03 models modified in this way indeed are a better match to the colours of the CMASS galaxies (Figs.~\ref{fig:paper2SEDfitcolcolbc03orig5} to \ref{fig:paper2SEDfitcolcolbc03orig8},  \ref{fig:paper2SEDfitcolcolbc03expbeta5} to \ref{fig:paper2SEDfitcolcolbc03expbeta8}, and \ref{fig:dcol5} to \ref{fig:dcol8}) and also yield better $\chi^2$ values from the fitting (Fig.~\ref{fig:chi2distr}).\\
From these best fitting \glspl{SED} we selected a small subset that should cover the region in colour space and absolute magnitude in $r$ of the CMASS sample.
We therefore projected the CMASS galaxy colours and $M_r$ of the four $z$ bins onto two-dimensional planes using a self-organizing map.
Afterwards we partition the plane in six clusters for each redshift bin with a $k$-means clustering algorithm and select one model \gls{SED} per cluster cell that produces the best \glspl{photoz} for galaxies within the same cell.
We estimated photometric redshifts with a template fitting code and with the selected models and analyse their individual performance.
Thereby we modified their redshift priors to improve on the \glspl{photoz}, but also decided to omit some of the templates which do not yield accurate \glspl{photoz} on the whole sample, regardless how the priors are modified.
We then compared the photometric redshift results with the \glspl{photoz} of two empirical methods published by \gls{SDSS} and calculated several metrics that assess the quality of the \glspl{photoz}, their estimated errors, and their distribution.
We found that the \glspl{photoz} with the generated models of this work produce better values in all quality metrics.
Furthermore, we observed that including the stacked \glspl{PDF} yield better results in the reconstruction of $N(\mzs)$ mirrored in the results of a KS test.
Concerning the estimated errors $\delta z$, a significance test shows that none of the three considered $\delta z$ results provides a reliable classification of outliers.
However, the probabilities of a non-outlier being misclassified is very small for all three considered cases.\\
Finally, we compared the predicted colours of the novel model \glspl{SED} when fitted to the \glspl{photoz} of this work and to the better of the two \gls{SDSS} redshifts, to the data.
We found that the deviations from the data are smaller for \glspl{photoz} of this work.\\

The models of this work and G13 can be downloaded together with the priors at http://www.cefca.es/people/\textasciitilde ngreisel/.
We would be happy to provide the photometric redshifts on request.

\subsubsection*{Acknowledgements}
We would like to thank the referee for her/his suggestions, which improved the manuscript.
We also thank Achim Weiss for discussions about \gls{AGB} modelling.\\

This work was supported by SFB-Transregio 33 (TR33) ``The Dark Universe'' by the Deutsche Forschungsgemeinschaft (DFG) and the DFG cluster of excellence ``Origin and Structure of the Universe``.\\

Funding for \gls{SDSS}-III has been provided by the Alfred P. Sloan Foundation, the Participating Institutions, the National Science Foundation, and the U.S. Department of Energy Office of Science.
The \gls{SDSS}-III web site is http://www.sdss3.org/.
\gls{SDSS}-III is managed by the Astrophysical Research Consortium for the Participating Institutions of the \gls{SDSS}-III Collaboration including the University of Arizona, the Brazilian Participation Group, Brookhaven National Laboratory, Carnegie Mellon University, University of Florida, the French Participation Group, the German Participation Group, Harvard University, the Instituto de Astrofisica de Canarias, the Michigan State/Notre Dame/JINA Participation Group, Johns Hopkins University, Lawrence Berkeley National Laboratory, Max Planck Institute for Astrophysics, Max Planck Institute for Extraterrestrial Physics, New Mexico State University, New York University, Ohio State University, Pennsylvania State University, University of Portsmouth, Princeton University, the Spanish Participation Group, University of Tokyo, University of Utah, Vanderbilt University, University of Virginia, University of Washington, and Yale University.

\bibliographystyle{mn2efix}
\bibliography{bibliography}

\begin{thebibliography}{72}
\expandafter\ifx\csname natexlab\endcsname\relax\def\natexlab#1{#1}\fi

\bibitem[{{Abazajian} {et~al}\mbox{.}(2009){Abazajian}, {Adelman-McCarthy},
  {Ag{\"u}eros}, {Allam}, {Allende Prieto}, {An}, {Anderson}, {Anderson},
  {Annis}, {Bahcall}, \& et~al.}]{2009ApJS..182..543A}
{Abazajian} K.~N. {et~al.}, 2009, ApJS, 182, 543

\bibitem[{{Ahn} {et~al}\mbox{.}(2014){Ahn}, {Alexandroff}, {Allende Prieto},
  {Anders}, {Anderson}, {Anderton}, {Andrews}, {Aubourg}, {Bailey}, {Bastien},
  \& et~al.}]{2014ApJS..211...17A}
{Ahn} C.~P. {et~al.}, 2014, ApJS, 211, 17

\bibitem[{{Arnouts} {et~al}\mbox{.}(1999){Arnouts}, {Cristiani}, {Moscardini},
  {Matarrese}, {Lucchin}, {Fontana}, \& {Giallongo}}]{1999MNRAS.310..540A}
{Arnouts} S., {Cristiani} S., {Moscardini} L., {Matarrese} S., {Lucchin} F.,
  {Fontana} A., {Giallongo} E., 1999, MNRAS, 310, 540

\bibitem[{{Baldry} {et~al}\mbox{.}(2004){Baldry}, {Glazebrook}, {Brinkmann},
  {Ivezi{\'c}}, {Lupton}, {Nichol}, \& {Szalay}}]{2004ApJ...600..681B}
{Baldry} I.~K., {Glazebrook} K., {Brinkmann} J., {Ivezi{\'c}} {\v Z}., {Lupton}
  R.~H., {Nichol} R.~C., {Szalay} A.~S., 2004, ApJ, 600, 681

\bibitem[{{Bender} {et~al}\mbox{.}(2001){Bender}, {Appenzeller}, {B{\"o}hm},
  {Drory}, {Fricke}, {Gabasch}, {Heidt}, {Hopp}, {J{\"a}ger}, {K{\"u}mmel},
  {Mehlert}, {M{\"o}llenhoff}, {Moorwood}, {Nicklas}, {Noll}, {Saglia},
  {Seifert}, {Seitz}, {Stahl}, {Sutorius}, {Szeifert}, {Wagner}, \&
  {Ziegler}}]{2001defi.conf...96B}
{Bender} R. {et~al.}, 2001, in Deep Fields, {S.~Cristiani, A.~Renzini, \&
  R.~E.~Williams}, ed., p.~96

\bibitem[{{Ben{\'{\i}}tez}(2000)}]{2000ApJ...536..571B}
{Ben{\'{\i}}tez} N., 2000, ApJ, 536, 571

\bibitem[{{Ben{\'{\i}}tez} {et~al}\mbox{.}(2009){Ben{\'{\i}}tez},
  {Gazta{\~n}aga}, {Miquel}, {Castander}, {Moles}, {Crocce},
  {Fern{\'a}ndez-Soto}, {Fosalba}, {Ballesteros}, {Campa}, {Cardiel-Sas},
  {Castilla}, {Crist{\'o}bal-Hornillos}, {Delfino}, {Fern{\'a}ndez},
  {Fern{\'a}ndez-Sopuerta}, {Garc{\'{\i}}a-Bellido}, {Lobo}, {Mart{\'{\i}}nez},
  {Ortiz}, {Pacheco}, {Paredes}, {Pons-Border{\'{\i}}a}, {S{\'a}nchez},
  {S{\'a}nchez}, {Varela}, \& {de Vicente}}]{2009ApJ...691..241B}
{Ben{\'{\i}}tez} N. {et~al.}, 2009, ApJ, 691, 241

\bibitem[{{Bolzonella}, {Miralles} \& {Pell{\'o}}(2000){Bolzonella},
  {Miralles}, \& {Pell{\'o}}}]{2000A&A...363..476B}
{Bolzonella} M., {Miralles} J.-M., {Pell{\'o}} R., 2000, A\&A, 363, 476

\bibitem[{{Brammer}, {van Dokkum} \& {Coppi}(2008){Brammer}, {van Dokkum}, \&
  {Coppi}}]{2008ApJ...686.1503B}
{Brammer} G.~B., {van Dokkum} P.~G., {Coppi} P., 2008, ApJ, 686, 1503

\bibitem[{{Brimioulle} {et~al}\mbox{.}(2008){Brimioulle}, {Lerchster}, {Seitz},
  {Bender}, \& {Snigula}}]{2008arXiv0811.3211B}
{Brimioulle} F., {Lerchster} M., {Seitz} S., {Bender} R., {Snigula} J., 2008,
  ArXiv e-prints

\bibitem[{{Brimioulle} {et~al}\mbox{.}(2013){Brimioulle}, {Seitz}, {Lerchster},
  {Bender}, \& {Snigula}}]{2013MNRAS.432.1046B}
{Brimioulle} F., {Seitz} S., {Lerchster} M., {Bender} R., {Snigula} J., 2013,
  MNRAS, 432, 1046

\bibitem[{{Bruzual} \& {Charlot}(2003)}]{2003MNRAS.344.1000B}
{Bruzual} G., {Charlot} S., 2003, MNRAS, 344, 1000

\bibitem[{{Calzetti} {et~al}\mbox{.}(2000){Calzetti}, {Armus}, {Bohlin},
  {Kinney}, {Koornneef}, \& {Storchi-Bergmann}}]{2000ApJ...533..682C}
{Calzetti} D., {Armus} L., {Bohlin} R.~C., {Kinney} A.~L., {Koornneef} J.,
  {Storchi-Bergmann} T., 2000, ApJ, 533, 682

\bibitem[{{Carliles} {et~al}\mbox{.}(2010){Carliles}, {Budav{\'a}ri}, {Heinis},
  {Priebe}, \& {Szalay}}]{2010ApJ...712..511C}
{Carliles} S., {Budav{\'a}ri} T., {Heinis} S., {Priebe} C., {Szalay} A.~S.,
  2010, ApJ, 712, 511

\bibitem[{{Carrasco Kind} \& {Brunner}(2013)}]{2013MNRAS.432.1483C}
{Carrasco Kind} M., {Brunner} R.~J., 2013, MNRAS, 432, 1483

\bibitem[{{Carrasco Kind} \& {Brunner}(2014)}]{2014MNRAS.438.3409C}
{Carrasco Kind} M., {Brunner} R.~J., 2014, MNRAS, 438, 3409

\bibitem[{{Cassar{\`a}} {et~al}\mbox{.}(2013){Cassar{\`a}}, {Piovan}, {Weiss},
  {Salaris}, \& {Chiosi}}]{2013MNRAS.436.2824C}
{Cassar{\`a}} L.~P., {Piovan} L., {Weiss} A., {Salaris} M., {Chiosi} C., 2013,
  MNRAS, 436, 2824

\bibitem[{{Cenarro} {et~al}\mbox{.}(2003){Cenarro}, {Gorgas}, {Vazdekis},
  {Cardiel}, \& {Peletier}}]{2003MNRAS.339L..12C}
{Cenarro} A.~J., {Gorgas} J., {Vazdekis} A., {Cardiel} N., {Peletier} R.~F.,
  2003, MNRAS, 339, L12

\bibitem[{{Chabrier}(2003)}]{2003PASP..115..763C}
{Chabrier} G., 2003, PASP, 115, 763

\bibitem[{{Collister} \& {Lahav}(2004)}]{2004PASP..116..345C}
{Collister} A.~A., {Lahav} O., 2004, PASP, 116, 345

\bibitem[{{Conroy} \& {van Dokkum}(2012)}]{2012ApJ...760...71C}
{Conroy} C., {van Dokkum} P.~G., 2012, ApJ, 760, 71

\bibitem[{{Csabai} {et~al}\mbox{.}(2003){Csabai}, {Budav{\'a}ri}, {Connolly},
  {Szalay}, {Gy{\H o}ry}, {Ben{\'{\i}}tez}, {Annis}, {Brinkmann}, {Eisenstein},
  {Fukugita}, {Gunn}, {Kent}, {Lupton}, {Nichol}, \&
  {Stoughton}}]{2003AJ....125..580C}
{Csabai} I. {et~al.}, 2003, AJ, 125, 580

\bibitem[{{Csabai} {et~al}\mbox{.}(2007){Csabai}, {Dobos}, {Trencs{\'e}ni},
  {Herczegh}, {J{\'o}zsa}, {Purger}, {Budav{\'a}ri}, \&
  {Szalay}}]{2007AN....328..852C}
{Csabai} I., {Dobos} L., {Trencs{\'e}ni} M., {Herczegh} G., {J{\'o}zsa} P.,
  {Purger} N., {Budav{\'a}ri} T., {Szalay} A.~S., 2007, Astronomische
  Nachrichten, 328, 852

\bibitem[{{Dawson} {et~al}\mbox{.}(2013){Dawson}, {Schlegel}, {Ahn},
  {Anderson}, {Aubourg}, {Bailey}, {Barkhouser}, {Bautista}, {Beifiori},
  {Berlind}, {Bhardwaj}, {Bizyaev}, {Blake}, {Blanton}, {Blomqvist}, {Bolton},
  {Borde}, {Bovy}, {Brandt}, {Brewington}, {Brinkmann}, {Brown}, {Brownstein},
  {Bundy}, {Busca}, {Carithers}, {Carnero}, {Carr}, {Chen}, {Comparat},
  {Connolly}, {Cope}, {Croft}, {Cuesta}, {da Costa}, {Davenport}, {Delubac},
  {de Putter}, {Dhital}, {Ealet}, {Ebelke}, {Eisenstein}, {Escoffier}, {Fan},
  {Filiz Ak}, {Finley}, {Font-Ribera}, {G{\'e}nova-Santos}, {Gunn}, {Guo},
  {Haggard}, {Hall}, {Hamilton}, {Harris}, {Harris}, {Ho}, {Hogg}, {Holder},
  {Honscheid}, {Huehnerhoff}, {Jordan}, {Jordan}, {Kauffmann}, {Kazin},
  {Kirkby}, {Klaene}, {Kneib}, {Le Goff}, {Lee}, {Long}, {Loomis}, {Lundgren},
  {Lupton}, {Maia}, {Makler}, {Malanushenko}, {Malanushenko}, {Mandelbaum},
  {Manera}, {Maraston}, {Margala}, {Masters}, {McBride}, {McDonald}, {McGreer},
  {McMahon}, {Mena}, {Miralda-Escud{\'e}}, {Montero-Dorta}, {Montesano},
  {Muna}, {Myers}, {Naugle}, {Nichol}, {Noterdaeme}, {Nuza}, {Olmstead},
  {Oravetz}, {Oravetz}, {Owen}, {Padmanabhan}, {Palanque-Delabrouille}, {Pan},
  {Parejko}, {P{\^a}ris}, {Percival}, {P{\'e}rez-Fournon},
  {P{\'e}rez-R{\`a}fols}, {Petitjean}, {Pfaffenberger}, {Pforr}, {Pieri},
  {Prada}, {Price-Whelan}, {Raddick}, {Rebolo}, {Rich}, {Richards}, {Rockosi},
  {Roe}, {Ross}, {Ross}, {Rossi}, {Rubi{\~n}o-Martin}, {Samushia},
  {S{\'a}nchez}, {Sayres}, {Schmidt}, {Schneider}, {Sc{\'o}ccola}, {Seo},
  {Shelden}, {Sheldon}, {Shen}, {Shu}, {Slosar}, {Smee}, {Snedden}, {Stauffer},
  {Steele}, {Strauss}, {Streblyanska}, {Suzuki}, {Swanson}, {Tal}, {Tanaka},
  {Thomas}, {Tinker}, {Tojeiro}, {Tremonti}, {Vargas Maga{\~n}a}, {Verde},
  {Viel}, {Wake}, {Watson}, {Weaver}, {Weinberg}, {Weiner}, {West}, {White},
  {Wood-Vasey}, {Yeche}, {Zehavi}, {Zhao}, \& {Zheng}}]{2013AJ....145...10D}
{Dawson} K.~S. {et~al.}, 2013, AJ, 145, 10

\bibitem[{{Drory}, {Bender} \& {Hopp}(2004){Drory}, {Bender}, \&
  {Hopp}}]{2004ApJ...616L.103D}
{Drory} N., {Bender} R., {Hopp} U., 2004, ApJ, 616, L103

\bibitem[{{Drory} {et~al}\mbox{.}(2001){Drory}, {Feulner}, {Bender}, {Botzler},
  {Hopp}, {Maraston}, {Mendes de Oliveira}, \& {Snigula}}]{2001MNRAS.325..550D}
{Drory} N., {Feulner} G., {Bender} R., {Botzler} C.~S., {Hopp} U., {Maraston}
  C., {Mendes de Oliveira} C., {Snigula} J., 2001, MNRAS, 325, 550

\bibitem[{{Drory} {et~al}\mbox{.}(2005){Drory}, {Salvato}, {Gabasch}, {Bender},
  {Hopp}, {Feulner}, \& {Pannella}}]{2005ApJ...619L.131D}
{Drory} N., {Salvato} M., {Gabasch} A., {Bender} R., {Hopp} U., {Feulner} G.,
  {Pannella} M., 2005, ApJ, 619, L131

\bibitem[{{Eisenstein} {et~al}\mbox{.}(2001){Eisenstein}, {Annis}, {Gunn},
  {Szalay}, {Connolly}, {Nichol}, {Bahcall}, {Bernardi}, {Burles}, {Castander},
  {Fukugita}, {Hogg}, {Ivezi{\'c}}, {Knapp}, {Lupton}, {Narayanan}, {Postman},
  {Reichart}, {Richmond}, {Schneider}, {Schlegel}, {Strauss}, {SubbaRao},
  {Tucker}, {Vanden Berk}, {Vogeley}, {Weinberg}, \&
  {Yanny}}]{2001AJ....122.2267E}
{Eisenstein} D.~J. {et~al.}, 2001, AJ, 122, 2267

\bibitem[{{Eisenstein} {et~al}\mbox{.}(2011){Eisenstein}, {Weinberg}, {Agol},
  {Aihara}, {Allende Prieto}, {Anderson}, {Arns}, {Aubourg}, {Bailey},
  {Balbinot}, \& et~al.}]{2011AJ....142...72E}
{Eisenstein} D.~J. {et~al.}, 2011, AJ, 142, 72

\bibitem[{{Feldmann} {et~al}\mbox{.}(2006){Feldmann}, {Carollo}, {Porciani},
  {Lilly}, {Capak}, {Taniguchi}, {Le F{\`e}vre}, {Renzini}, {Scoville},
  {Ajiki}, {Aussel}, {Contini}, {McCracken}, {Mobasher}, {Murayama}, {Sanders},
  {Sasaki}, {Scarlata}, {Scodeggio}, {Shioya}, {Silverman}, {Takahashi},
  {Thompson}, \& {Zamorani}}]{2006MNRAS.372..565F}
{Feldmann} R. {et~al.}, 2006, MNRAS, 372, 565

\bibitem[{{Feulner} {et~al}\mbox{.}(2005){Feulner}, {Gabasch}, {Salvato},
  {Drory}, {Hopp}, \& {Bender}}]{2005ApJ...633L...9F}
{Feulner} G., {Gabasch} A., {Salvato} M., {Drory} N., {Hopp} U., {Bender} R.,
  2005, ApJ, 633, L9

\bibitem[{{Gabasch} {et~al}\mbox{.}(2004){Gabasch}, {Bender}, {Seitz}, {Hopp},
  {Saglia}, {Feulner}, {Snigula}, {Drory}, {Appenzeller}, {Heidt}, {Mehlert},
  {Noll}, {B{\"o}hm}, {J{\"a}ger}, {Ziegler}, \&
  {Fricke}}]{2004A&A...421...41G}
{Gabasch} A. {et~al.}, 2004, A\&A, 421, 41

\bibitem[{{Gabasch} {et~al}\mbox{.}(2008){Gabasch}, {Goranova}, {Hopp}, {Noll},
  \& {Pannella}}]{2008MNRAS.383.1319G}
{Gabasch} A., {Goranova} Y., {Hopp} U., {Noll} S., {Pannella} M., 2008, MNRAS,
  383, 1319

\bibitem[{{Gerdes}(2009)}]{2009AAS...21348303G}
{Gerdes} D.~W., 2009, in Bulletin of the American Astronomical Society,
  Vol.~41, American Astronomical Society Meeting Abstracts \#213, p. \#483.03

\bibitem[{{Greisel} {et~al}\mbox{.}(2013){Greisel}, {Seitz}, {Drory}, {Bender},
  {Saglia}, \& {Snigula}}]{2013ApJ...768..117G}
{Greisel} N., {Seitz} S., {Drory} N., {Bender} R., {Saglia} R.~P., {Snigula}
  J., 2013, ApJ, 768, 117

\bibitem[{{Gruen} {et~al}\mbox{.}(2013){Gruen}, {Brimioulle}, {Seitz}, {Lee},
  {Young}, {Koppenhoefer}, {Eichner}, {Riffeser}, {Vikram}, {Weidinger}, \&
  {Zenteno}}]{2013MNRAS.432.1455G}
{Gruen} D. {et~al.}, 2013, MNRAS, 432, 1455

\bibitem[{{Gruen} {et~al}\mbox{.}(2014){Gruen}, {Seitz}, {Brimioulle},
  {Kosyra}, {Koppenhoefer}, {Lee}, {Bender}, {Riffeser}, {Eichner},
  {Weidinger}, \& {Bierschenk}}]{2014MNRAS.442.1507G}
{Gruen} D. {et~al.}, 2014, MNRAS, 442, 1507

\bibitem[{{Hanke} {et~al}\mbox{.}(2009){Hanke}, {Halchenko}, {Sederberg},
  {Hanson}, {Haxby}, \& {Pollmann}}]{PyMVPA}
{Hanke} M., {Halchenko} Y.~O., {Sederberg} P.~B., {Hanson} S.~J., {Haxby}
  J.~V., {Pollmann} S., 2009, Neuroinformatics, 7, 37

\bibitem[{{Hildebrandt} {et~al}\mbox{.}(2012){Hildebrandt}, {Erben}, {Kuijken},
  {van Waerbeke}, {Heymans}, {Coupon}, {Benjamin}, {Bonnett}, {Fu}, {Hoekstra},
  {Kitching}, {Mellier}, {Miller}, {Velander}, {Hudson}, {Rowe}, {Schrabback},
  {Semboloni}, \& {Ben{\'{\i}}tez}}]{2012MNRAS.421.2355H}
{Hildebrandt} H. {et~al.}, 2012, MNRAS, 421, 2355

\bibitem[{{Huterer} {et~al}\mbox{.}(2006){Huterer}, {Takada}, {Bernstein}, \&
  {Jain}}]{2006MNRAS.366..101H}
{Huterer} D., {Takada} M., {Bernstein} G., {Jain} B., 2006, MNRAS, 366, 101

\bibitem[{{Ilbert} {et~al}\mbox{.}(2006){Ilbert}, {Arnouts}, {McCracken},
  {Bolzonella}, {Bertin}, {Le F{\`e}vre}, {Mellier}, {Zamorani}, {Pell{\`o}},
  {Iovino}, {Tresse}, {Le Brun}, {Bottini}, {Garilli}, {Maccagni}, {Picat},
  {Scaramella}, {Scodeggio}, {Vettolani}, {Zanichelli}, {Adami}, {Bardelli},
  {Cappi}, {Charlot}, {Ciliegi}, {Contini}, {Cucciati}, {Foucaud}, {Franzetti},
  {Gavignaud}, {Guzzo}, {Marano}, {Marinoni}, {Mazure}, {Meneux}, {Merighi},
  {Paltani}, {Pollo}, {Pozzetti}, {Radovich}, {Zucca}, {Bondi}, {Bongiorno},
  {Busarello}, {de La Torre}, {Gregorini}, {Lamareille}, {Mathez}, {Merluzzi},
  {Ripepi}, {Rizzo}, \& {Vergani}}]{2006A&A...457..841I}
{Ilbert} O. {et~al.}, 2006, A\&A, 457, 841

\bibitem[{Kohonen(1982)}]{SOM}
Kohonen T., 1982, Biological Cybernetics, 43, 59

\bibitem[{{Kohonen}(2001)}]{2001som..book.....K}
{Kohonen} T., 2001, {Self-Organizing Maps}

\bibitem[{{Kroupa}(2001)}]{2001MNRAS.322..231K}
{Kroupa} P., 2001, MNRAS, 322, 231

\bibitem[{{Longhetti} \& {Saracco}(2009)}]{2009MNRAS.394..774L}
{Longhetti} M., {Saracco} P., 2009, MNRAS, 394, 774

\bibitem[{{Ma}, {Hu} \& {Huterer}(2006){Ma}, {Hu}, \&
  {Huterer}}]{2006ApJ...636...21M}
{Ma} Z., {Hu} W., {Huterer} D., 2006, ApJ, 636, 21

\bibitem[{{Mancone} \& {Gonzalez}(2012)}]{2012PASP..124..606M}
{Mancone} C.~L., {Gonzalez} A.~H., 2012, PASP, 124, 606

\bibitem[{{Mandelbaum} {et~al}\mbox{.}(2008){Mandelbaum}, {Seljak}, {Hirata},
  {Bardelli}, {Bolzonella}, {Bongiorno}, {Carollo}, {Contini}, {Cunha},
  {Garilli}, {Iovino}, {Kampczyk}, {Kneib}, {Knobel}, {Koo}, {Lamareille}, {Le
  F{\`e}vre}, {Le Borgne}, {Lilly}, {Maier}, {Mainieri}, {Mignoli}, {Newman},
  {Oesch}, {Perez-Montero}, {Ricciardelli}, {Scodeggio}, {Silverman}, \&
  {Tasca}}]{2008MNRAS.386..781M}
{Mandelbaum} R. {et~al.}, 2008, MNRAS, 386, 781

\bibitem[{{Maraston}(1998)}]{1998MNRAS.300..872M}
{Maraston} C., 1998, MNRAS, 300, 872

\bibitem[{{Maraston}(2005)}]{2005MNRAS.362..799M}
{Maraston} C., 2005, MNRAS, 362, 799

\bibitem[{{Maraston} {et~al}\mbox{.}(2010){Maraston}, {Pforr}, {Renzini},
  {Daddi}, {Dickinson}, {Cimatti}, \& {Tonini}}]{2010MNRAS.407..830M}
{Maraston} C., {Pforr} J., {Renzini} A., {Daddi} E., {Dickinson} M., {Cimatti}
  A., {Tonini} C., 2010, MNRAS, 407, 830

\bibitem[{{Maraston} \& {Str{\"o}mb{\"a}ck}(2011)}]{2011MNRAS.418.2785M}
{Maraston} C., {Str{\"o}mb{\"a}ck} G., 2011, MNRAS, 418, 2785

\bibitem[{{Maraston} {et~al}\mbox{.}(2009){Maraston}, {Str{\"o}mb{\"a}ck},
  {Thomas}, {Wake}, \& {Nichol}}]{2009MNRAS.394L.107M}
{Maraston} C., {Str{\"o}mb{\"a}ck} G., {Thomas} D., {Wake} D.~A., {Nichol}
  R.~C., 2009, MNRAS, 394, L107

\bibitem[{{Marigo} {et~al}\mbox{.}(2008){Marigo}, {Girardi}, {Bressan},
  {Groenewegen}, {Silva}, \& {Granato}}]{2008A&A...482..883M}
{Marigo} P., {Girardi} L., {Bressan} A., {Groenewegen} M.~A.~T., {Silva} L.,
  {Granato} G.~L., 2008, A\&A, 482, 883

\bibitem[{{Meidt} {et~al}\mbox{.}(2012){Meidt}, {Schinnerer}, {Knapen},
  {Bosma}, {Athanassoula}, {Sheth}, {Buta}, {Zaritsky}, {Laurikainen},
  {Elmegreen}, {Elmegreen}, {Gadotti}, {Salo}, {Regan}, {Ho}, {Madore}, {Hinz},
  {Skibba}, {Gil de Paz}, {Mu{\~n}oz-Mateos}, {Men{\'e}ndez-Delmestre},
  {Seibert}, {Kim}, {Mizusawa}, {Laine}, \&
  {Comer{\'o}n}}]{2012ApJ...744...17M}
{Meidt} S.~E. {et~al.}, 2012, ApJ, 744, 17

\bibitem[{{Monna} {et~al}\mbox{.}(2014){Monna}, {Seitz}, {Greisel}, {Eichner},
  {Drory}, {Postman}, {Zitrin}, {Coe}, {Halkola}, {Suyu}, {Grillo}, {Rosati},
  {Lemze}, {Balestra}, {Snigula}, {Bradley}, {Umetsu}, {Koekemoer}, {Kuchner},
  {Moustakas}, {Bartelmann}, {Ben{\'{\i}}tez}, {Bouwens}, {Broadhurst},
  {Donahue}, {Ford}, {Host}, {Infante}, {Jimenez-Teja}, {Jouvel}, {Kelson},
  {Lahav}, {Medezinski}, {Melchior}, {Meneghetti}, {Merten}, {Molino},
  {Moustakas}, {Nonino}, \& {Zheng}}]{2014MNRAS.438.1417M}
{Monna} A. {et~al.}, 2014, MNRAS, 438, 1417

\bibitem[{{Oke} \& {Gunn}(1983)}]{1983ApJ...266..713O}
{Oke} J.~B., {Gunn} J.~E., 1983, ApJ, 266, 713

\bibitem[{{Padmanabhan} {et~al}\mbox{.}(2005){Padmanabhan}, {Budav{\'a}ri},
  {Schlegel}, {Bridges}, {Brinkmann}, {Cannon}, {Connolly}, {Croom}, {Csabai},
  {Drinkwater}, {Eisenstein}, {Hewett}, {Loveday}, {Nichol}, {Pimbblet}, {De
  Propris}, {Schneider}, {Scranton}, {Seljak}, {Shanks}, {Szapudi}, {Szalay},
  \& {Wake}}]{2005MNRAS.359..237P}
{Padmanabhan} N. {et~al.}, 2005, MNRAS, 359, 237

\bibitem[{Pedregosa {et~al}\mbox{.}(2011)Pedregosa, Varoquaux, Gramfort,
  Michel, Thirion, Grisel, Blondel, Prettenhofer, Weiss, Dubourg, Vanderplas,
  Passos, Cournapeau, Brucher, Perrot, \& Duchesnay}]{scikit-learn}
Pedregosa F. {et~al.}, 2011, Journal of Machine Learning Research, 12, 2825

\bibitem[{{Pforr}, {Maraston} \& {Tonini}(2012){Pforr}, {Maraston}, \&
  {Tonini}}]{2012MNRAS.422.3285P}
{Pforr} J., {Maraston} C., {Tonini} C., 2012, MNRAS, 422, 3285

\bibitem[{{Pickles}(1998)}]{1998PASP..110..863P}
{Pickles} A.~J., 1998, PASP, 110, 863

\bibitem[{{Saglia} {et~al}\mbox{.}(2012){Saglia}, {Tonry}, {Bender}, {Greisel},
  {Seitz}, {Senger}, {Snigula}, {Phleps}, {Wilman}, {Bailer-Jones}, {Klement},
  {Rix}, {Smith}, {Green}, {Burgett}, {Chambers}, {Heasley}, {Kaiser},
  {Magnier}, {Morgan}, {Price}, {Stubbs}, \& {Wainscoat}}]{2012ApJ...746..128S}
{Saglia} R.~P. {et~al.}, 2012, ApJ, 746, 128

\bibitem[{{Salaris} {et~al}\mbox{.}(2014){Salaris}, {Weiss}, {Cassar{\`a}},
  {Piovan}, \& {Chiosi}}]{2014A&A...565A...9S}
{Salaris} M., {Weiss} A., {Cassar{\`a}} L.~P., {Piovan} L., {Chiosi} C., 2014,
  A\&A, 565, A9

\bibitem[{{Salpeter}(1955)}]{1955ApJ...121..161S}
{Salpeter} E.~E., 1955, ApJ, 121, 161

\bibitem[{{S{\'a}nchez} {et~al}\mbox{.}(2014){S{\'a}nchez}, {Carrasco Kind},
  {Lin}, {Miquel}, {Abdalla}, {Amara}, {Banerji}, {Bonnett}, {Brunner},
  {Capozzi}, {Carnero}, {Castander}, {da Costa}, {Cunha}, {Fausti}, {Gerdes},
  {Greisel}, {Gschwend}, {Hartley}, {Jouvel}, {Lahav}, {Lima}, {Maia},
  {Mart{\'{\i}}}, {Ogando}, {Ostrovski}, {Pellegrini}, {Rau}, {Sadeh}, {Seitz},
  {Sevilla-Noarbe}, {Sypniewski}, {de Vicente}, {Abbot}, {Allam}, {Atlee},
  {Bernstein}, {Bernstein}, {Buckley-Geer}, {Burke}, {Childress}, {Davis},
  {DePoy}, {Dey}, {Desai}, {Diehl}, {Doel}, {Estrada}, {Evrard},
  {Fern{\'a}ndez}, {Finley}, {Flaugher}, {Frieman}, {Gaztanaga}, {Glazebrook},
  {Honscheid}, {Kim}, {Kuehn}, {Kuropatkin}, {Lidman}, {Makler}, {Marshall},
  {Nichol}, {Roodman}, {S{\'a}nchez}, {Santiago}, {Sako}, {Scalzo}, {Smith},
  {Swanson}, {Tarle}, {Thomas}, {Tucker}, {Uddin}, {Vald{\'e}s}, {Walker},
  {Yuan}, \& {Zuntz}}]{2014MNRAS.445.1482S}
{S{\'a}nchez} C. {et~al.}, 2014, MNRAS, 445, 1482

\bibitem[{{S{\'a}nchez} {et~al}\mbox{.}(2011){S{\'a}nchez}, {Carnero},
  {Garc{\'{\i}}a-Bellido}, {Gazta{\~n}aga}, {de Simoni}, {Crocce}, {Cabr{\'e}},
  {Fosalba}, \& {Alonso}}]{2011MNRAS.411..277S}
{S{\'a}nchez} E. {et~al.}, 2011, MNRAS, 411, 277

\bibitem[{{Shapley} {et~al}\mbox{.}(2005){Shapley}, {Steidel}, {Erb}, {Reddy},
  {Adelberger}, {Pettini}, {Barmby}, \& {Huang}}]{2005ApJ...626..698S}
{Shapley} A.~E., {Steidel} C.~C., {Erb} D.~K., {Reddy} N.~A., {Adelberger}
  K.~L., {Pettini} M., {Barmby} P., {Huang} J., 2005, ApJ, 626, 698

\bibitem[{{Smee} {et~al}\mbox{.}(2013){Smee}, {Gunn}, {Uomoto}, {Roe},
  {Schlegel}, {Rockosi}, {Carr}, {Leger}, {Dawson}, {Olmstead}, {Brinkmann},
  {Owen}, {Barkhouser}, {Honscheid}, {Harding}, {Long}, {Lupton}, {Loomis},
  {Anderson}, {Annis}, {Bernardi}, {Bhardwaj}, {Bizyaev}, {Bolton},
  {Brewington}, {Briggs}, {Burles}, {Burns}, {Castander}, {Connolly},
  {Davenport}, {Ebelke}, {Epps}, {Feldman}, {Friedman}, {Frieman}, {Heckman},
  {Hull}, {Knapp}, {Lawrence}, {Loveday}, {Mannery}, {Malanushenko},
  {Malanushenko}, {Merrelli}, {Muna}, {Newman}, {Nichol}, {Oravetz}, {Pan},
  {Pope}, {Ricketts}, {Shelden}, {Sandford}, {Siegmund}, {Simmons}, {Smith},
  {Snedden}, {Schneider}, {SubbaRao}, {Tremonti}, {Waddell}, \&
  {York}}]{2013AJ....146...32S}
{Smee} S.~A. {et~al.}, 2013, AJ, 146, 32

\bibitem[{{Steinhaus}(1957)}]{zbMATH03129892}
{Steinhaus} H., 1957, {Bull. Acad. Pol. Sci., Cl. III}, 4, 801

\bibitem[{{Voronoi}(1908)}]{voronoi}
{Voronoi} G., 1908, {Journal f\"ur die Reine und Angewandte Mathematik}, 133,
  97–178

\bibitem[{{York} {et~al}\mbox{.}(2000){York}, {Adelman}, {Anderson},
  {Anderson}, {Annis}, {Bahcall}, {Bakken}, {Barkhouser}, {Bastian}, {Berman},
  {Boroski}, {Bracker}, {Briegel}, {Briggs}, {Brinkmann}, {Brunner}, {Burles},
  {Carey}, {Carr}, {Castander}, {Chen}, {Colestock}, {Connolly}, {Crocker},
  {Csabai}, {Czarapata}, {Davis}, {Doi}, {Dombeck}, {Eisenstein}, {Ellman},
  {Elms}, {Evans}, {Fan}, {Federwitz}, {Fiscelli}, {Friedman}, {Frieman},
  {Fukugita}, {Gillespie}, {Gunn}, {Gurbani}, {de Haas}, {Haldeman}, {Harris},
  {Hayes}, {Heckman}, {Hennessy}, {Hindsley}, {Holm}, {Holmgren}, {Huang},
  {Hull}, {Husby}, {Ichikawa}, {Ichikawa}, {Ivezi{\'c}}, {Kent}, {Kim},
  {Kinney}, {Klaene}, {Kleinman}, {Kleinman}, {Knapp}, {Korienek}, {Kron},
  {Kunszt}, {Lamb}, {Lee}, {Leger}, {Limmongkol}, {Lindenmeyer}, {Long},
  {Loomis}, {Loveday}, {Lucinio}, {Lupton}, {MacKinnon}, {Mannery}, {Mantsch},
  {Margon}, {McGehee}, {McKay}, {Meiksin}, {Merelli}, {Monet}, {Munn},
  {Narayanan}, {Nash}, {Neilsen}, {Neswold}, {Newberg}, {Nichol}, {Nicinski},
  {Nonino}, {Okada}, {Okamura}, {Ostriker}, {Owen}, {Pauls}, {Peoples},
  {Peterson}, {Petravick}, {Pier}, {Pope}, {Pordes}, {Prosapio},
  {Rechenmacher}, {Quinn}, {Richards}, {Richmond}, {Rivetta}, {Rockosi},
  {Ruthmansdorfer}, {Sandford}, {Schlegel}, {Schneider}, {Sekiguchi}, {Sergey},
  {Shimasaku}, {Siegmund}, {Smee}, {Smith}, {Snedden}, {Stone}, {Stoughton},
  {Strauss}, {Stubbs}, {SubbaRao}, {Szalay}, {Szapudi}, {Szokoly}, {Thakar},
  {Tremonti}, {Tucker}, {Uomoto}, {Vanden Berk}, {Vogeley}, {Waddell}, {Wang},
  {Watanabe}, {Weinberg}, {Yanny}, {Yasuda}, \& {SDSS
  Collaboration}}]{2000AJ....120.1579Y}
{York} D.~G. {et~al.}, 2000, AJ, 120, 1579

\bibitem[{{Zibetti} {et~al}\mbox{.}(2013){Zibetti}, {Gallazzi}, {Charlot},
  {Pierini}, \& {Pasquali}}]{2013MNRAS.428.1479Z}
{Zibetti} S., {Gallazzi} A., {Charlot} S., {Pierini} D., {Pasquali} A., 2013,
  MNRAS, 428, 1479

\end{thebibliography}

\appendix
\section{SED Fitting Results}
\label{app:likelihood}
\begin{figure*}
  \centering
  \includegraphics[width=0.7\textwidth]{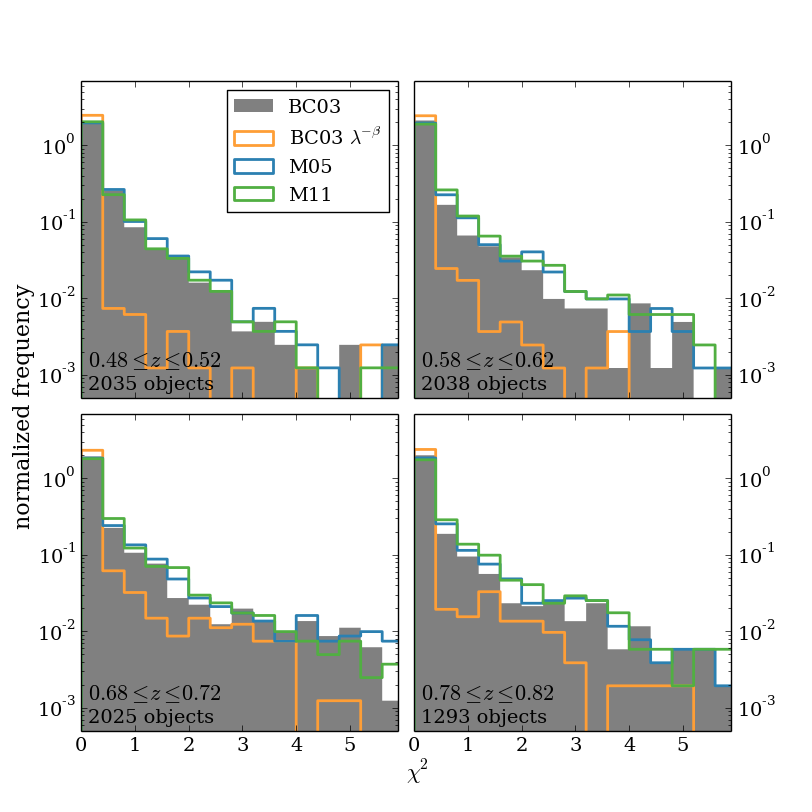}
  \caption{
    $\chi^2$ distributions of \gls{SED} fitting runs (at $\mzs$) detailed with BC03 (grey), \bcexpb\ (orange), M05 (blue), and M11 (green) models for the four considered redshift regions.
    The frequency is normalised and plotted logarithmically.
    We can see that although the distributions for BC03, M05, and M11 are similar, the $\chi^2$ values are improved by including the \bcexpb\ models.
  }
  \label{fig:chi2distr}
\end{figure*}

In this paper we create models that should match the \gls{BOSS} CMASS galaxies in terms of their colours in redshift bins with widths $0.04$ centred on $z=0.5,0.6,0.7,0.8$.
We do so by fitting sets of model \glspl{SED} with different properties like metallicity, \glspl{SFH}, and ages to the data (Sec.~\ref{sec:paper2SEDfittingresults}).
In superposition to the \glspl{SED}, a burst model and additional dust extinctions for the burst and the main component are further degrees of freedom.
In Fig.~\ref{fig:chi2distr} we show the distributions of the $\chi^2$ values returned by the \gls{SED} fitting code, when we fit the BC03, M05, and M11 models in equal setups to the data.
The setup of the \gls{SED} fitting and the code are detailed in Sec.~\ref{sec:paper2SEDfitting} and is equal for all cases.\\
In Sec.~\ref{sec:paper2SEDfittingresults} we modify the BC03 models, changing their continuum slope by multiplying with $a\cdot\lambda^{-\beta}$ for $\lambda\geq\lambda_i$ with various $\beta$ and $\lambda_i$ values.
This is done to introduce further degrees of freedom and to create models that match the colours of the data to a better extent.
We fit the \expb\ modifications of BC03 \glspl{SSP} and \glspl{CSP} with decreasing \glspl{SFR} to the CMASS galaxy colours with the same setup as mentioned above and as detailed in Sec.~\ref{sec:paper2SEDfittingresults}.
The resulting $\chi^2$ values of the fit are shown together with the previous results of the BC03, M05, and M11 models in Fig.~\ref{fig:chi2distr}, where they are highlighted by green histograms.\\
While the results of the original BC03 models, M05, and M11 produce similar results in terms of $\chi^2$, the fits are significantly improved by using the \bcexpb\ models.
This is also shown in terms of colours in Figs.~\ref{fig:dcol5} to \ref{fig:dcol8} of Sec.~\ref{sec:paper2SEDfittingresults}.

\section{Query for CMASS Sample}
\label{app:paper2query}
We present here the SQL query executed to download the data from the \gls{SDSS}-III \ttt{CasJobs} website\footnote{http://skyserver.sdss3.org/CasJobs/} used in this work and detailed in Sec.~\ref{sec:paper2BOSSCMASSsample}.
In the following \ttt{<fil>} denotes any of the five Sloan filters and can be replaced by \ttt{u}, \ttt{g}, \ttt{r}, \ttt{i}, and \ttt{z}.
\begin{verbatim}
select p.ObjID, s.z as zSpec,
       s.CModelMag_<fil>, s.CModelMagErr_<fil>, 
       s.Extinction_<fil>, 
       p.CModelFlux_<fil>, p.CModelFluxIvar_<fil>, 
       pz.z as zPhot, pz.zErr as zPhotErr, 
       pzrf.z as zPhotRF, pzrf.zErr as zPhotRFErr
from specPhotoall s, spa a, Photoobjall p,
     Photoz pz, PhotozRF pzrf
where s.SpecObjID=a.SpecObjID and
      s.ObjID=p.ObjID and pz.ObjID=p.ObjID and
      pzrf.ObjID=p.ObjID and
      s.zWarning=0 and s.z>0 and 
      (a.BOSS_TARGET1 & 2>0 or
      (a.BOSS_TARGET1 & 1>0 and s.Tile>=10324)) and
      a.BOSSPrimary=1 and a.zWarning_NoQSO=0 and 
      a.Chunk!='BOSS1' and a.Chunk!='BOSS2' and
      p.Fiber2Mag_i<21.5 and p.Clean=1 and
      (p.CalibStatus_<fil> & 1)!=0 and
      s.z>=0.45 and s.z <=0.9
\end{verbatim}

\label{lastpage}
\end{document}